\documentclass[journal, 10 pt]{IEEEtran}
\usepackage{amsmath,amsfonts}
\usepackage{algorithmic}
\usepackage{array}
\usepackage[caption=false,font=normalsize,labelfont=sf,textfont=sf]{subfig}
\usepackage{textcomp}
\usepackage{stfloats}
\usepackage{url}
\usepackage{verbatim}
\usepackage{amssymb}
\usepackage{xcolor}
\usepackage{graphicx}
\usepackage{balance}
\usepackage{multirow}

\begin{document}
	\title{Multi-Block Attention for Efficient Channel Estimation in IRS-Assisted mmWave MIMO}
	
	\author{Mehrdad Momen-Tayefeh$^{1}$, Mehrshad Momen-Tayefeh$^{2}$, Maryam Sabbaghian$^{1*}$ 
		\thanks{$^{1}$Mehrdad Momen-Tayefeh and Maryam Sabbaghian are with the School of Electrical and Computer Engineering, University of Tehran, Tehran, Iran (e-mail: mehrdad.momen@ut.ac.ir; msabbaghian@ut.ac.ir). \\
			$^{2}$Mehrshad Momen-Tayefeh is with the Department of Computer Engineering, Sharif University of Technology, Tehran, Iran (e-mail: mehrshadmomen@sharif.edu). \\
			*Corresponding Author: Maryam Sabbaghian (e-mail: msabbaghian@ut.ac.ir). 
			2025 IEEE. Personal use of this material is permitted. Permission from IEEE must be obtained for all other uses, in any current or future media. This is the author’s version of the work. It is posted here for your personal use. Not for redistribution. The definitive Version of Record was published in IEEE Transactions on Communications, DOI: 10.1109/TCOMM.2025.3618696.}}

	{}
	
	\maketitle
	\begin{abstract}
	Intelligent Reflecting Surfaces (IRSs) are a promising technology for enhancing the spectral and energy efficiency of millimeter-wave (mmWave) multiple-input multiple-output (MIMO) systems. In these systems, accurate channel estimation remains challenging due to the passive nature of IRS elements and the high pilot overhead in large-scale deployments. This paper presents a deep learning-based Multi-Block Attention (MBA) framework for efficient cascaded channel estimation in IRS-assisted mmWave MIMO systems that utilize orthogonal frequency division multiplexing (OFDM). First, we show the optimality of the discrete Fourier transform (DFT) and Hadamard matrices as phase configurations for least squares (LS) estimation. To reduce training overhead, we selectively deactivate IRS elements and compensate for induced feature loss using a two-stage architecture: (i) a Convolutional Attention Network (CAN) for spatial correlation recovery and (ii) a Complex Multi-Convolutional Network (CMN) for noise suppression. The MBA architecture mitigates error propagation through attention-guided feature refinement and denoising. Simulation results indicate that the MBA method reduces pilot overhead by up to 87\% compared to the LS estimator. Additionally, at signal-to-noise ratios of 10 dB, our proposed method achieves approximately 51\% lower normalized mean squared error (NMSE) than leading methods. It also maintains low computational complexity and adapts effectively to various propagation environments.
	\end{abstract}
	
	\begin{IEEEkeywords}
		Intelligent reflecting surfaces, channel estimation, deep learning, MIMO, millimeter-wave.
	\end{IEEEkeywords}
	
	\section{Introduction}
	\IEEEPARstart{I}{ntelligent} reflecting surfaces (IRSs) have emerged as a promising solution for enhancing coverage, link quality, and throughput in future wireless communication systems \cite{9326394}. An IRS consists of a planar array of passive elements capable of dynamically controlling the phase of incident signals to reconfigure the wireless propagation environment and reflect signals toward desired directions \cite{9122596}. Unlike active relays, IRSs operate passively, offering energy-efficient and cost-efficient enhancements without requiring additional radio-frequency (RF) chains \cite{9119122, 8811733}.
	
	In IRS-assisted systems, accurate knowledge of the cascaded channel state information (CSI), representing the combined base station (BS)-IRS and IRS-mobile user (MU) links, is essential for configuring the IRS phase shifts to maximize signal alignment and system throughput \cite{9086766}. However, acquiring accurate CSI is particularly challenging due to the passive nature of IRSs and the lack of baseband processing capabilities. This issue is further exacerbated in IRS-assisted orthogonal frequency division multiplexing (OFDM) systems, where the cascade channels over multiple subcarriers must be estimated concurrently \cite{8937491}. Therefore, developing efficient channel estimation techniques that can accurately recover the cascade channel across subcarriers with minimal pilot overhead is crucial for implementing IRS-assisted millimeter-wave (mmWave) multiple-input multiple-output (MIMO) OFDM systems.
	
	\subsection{Prior Works}
	Channel estimation in IRS-assisted wireless systems has received considerable attention in recent years. However, existing approaches face persistent challenges in balancing estimation accuracy, training overhead, and computational efficiency. 
	
	Early IRS-assisted channel estimation methods relied on sequential estimation by activating IRS elements one at a time, as in \cite{8683663}. While simple, this approach introduces prohibitive training overhead due to the large number of IRS elements. To mitigate this, Jensen et al. proposed a sub-surface grouping strategy coupled with minimum variance unbiased estimation \cite{9053695}, later extended to frequency-selective channels in \cite{yang2020intelligent}.
	For wideband single-input single-output (SISO) systems, Zheng et al. employed the least squares (LS) method for estimating frequency-selective channels in IRS-aided OFDM scenarios \cite{9195133}. Alternatively, minimum mean square error (MMSE) estimators under Gaussian assumptions offer closed-form solutions but often lack robustness in practical environments \cite{alwazani2020intelligent}.
	
	Channel estimation methods utilize the inherent sparsity in both the angular and delay domains of mmWave channels to minimize pilot overhead. In their work, J. He et al. introduced atomic norm minimization techniques for the super-resolution estimation of angle-of-departure (AoD) and angle-of-arrival (AoA) in \cite{9398559}. This approach was later extended to OFDM systems in \cite{10521790}. Additionally, structured matrix recovery and compressed sensing (CS) methods have been proposed in \cite{10053657, 10077537}. However, these methods have high computational complexity, particularly when applied to large IRS deployments.
	
	To overcome these challenges, low-rank matrix completion methods for multi-user cascaded channel estimation were proposed in \cite{10403782}. Although effective in principle, this approach suffers from high sensitivity to inaccuracies in rank estimation, which can significantly degrade performance. In the wideband context, the authors in \cite{10529332} developed orthogonal matching pursuit (OMP) and SBL-based algorithms, which exploit joint time–angle sparsity for SIMO-OFDM systems. However, their computational complexity scales cubically with the number of IRS elements and quadratically with the number of BS antennas, limiting scalability.
	
	More recently, a low-rank sparse tensor recovery method was proposed in \cite{10904337} for IRS-assisted mmWave OFDM systems. This approach models the cascaded channel as a low-rank sparse tensor in the angle-delay domain and reconstructs it via manifold optimization. Although this reduces pilot overhead and improves accuracy, it requires accurate prior knowledge and remains computationally intensive. Moreover, most CS-based techniques involve solving iterative optimization problems, often requiring high-resolution dictionaries that significantly increase complexity and estimation time.
	
	Deep learning (DL) has emerged as a powerful alternative for IRS-aided channel estimation due to its data-driven modeling capability and potential for real-time inference \cite{eldar2022machine, 8755300, 8054694}. Xu et al. in \cite{9611281} proposed a DL model combining an ODE-enhanced recurrent neural network for temporal interpolation and an ODE-inspired feedforward network for spatial extrapolation to estimate time-varying IRS-assisted channels from partial pilot observations.
	
	To enhance estimation while reducing hardware dependency, Liu et al. \cite{liu2020deep} incorporated additional RF chains at the IRS and utilized denoising convolutional neural networks (DnCNNs) for estimation. Gong et al. \cite{10566602} expanded this idea by equipping IRS elements with ADCs and deploying a deep neural network (DNN) to infer channel states. While these methods improve accuracy, they introduce high hardware costs and complexity due to the use of active IRS components. Similarly, Zhang et al. \cite{9489307} proposed a two-stage DNN approach for channel reconstruction and beam selection, requiring active IRS elements.
	
	Mao et al. \cite{mao2022channel} presented RS-OMP, which integrates residual learning with OMP. This was further extended by Abdallah et al. in \cite{9944694} through DD-FF and DD-FS models, which exploit angular sparsity for both frequency-flat and frequency-selective channels. While effective in reducing pilot overhead, these methods degrade significantly when the number of pilots is much smaller than the number of IRS elements.
	
	In \cite{10025776}, a super-resolution DNN approach was proposed using cascaded SRCNN and DnCNN networks for IRS-assisted MIMO-OFDM systems. Though it achieves high estimation accuracy, it still requires many time slots equal to the number of IRS elements, limiting efficiency gains. In \cite{10309967}, hybrid-driven learning strategies were introduced for CSI acquisition in IRS-aided wideband systems. The authors proposed the DA-RLAMP network for passive IRS, which integrates model-driven RLAMP with denoising and attention mechanisms, and further extended it to hybrid IRS through the lightweight MDA-RLAMP network.
	
	Generative adversarial networks (GAN) based approaches have also been explored. For instance, the authors in \cite{10328721} introduced a GAN model for narrowband channel estimation. However, it struggles to reduce training overhead in large-scale scenarios and cannot be extended to frequency-selective systems. Similarly, \cite{9761227} used conditional GANs for narrowband fading to enhance the accuracy of the cascade channel.
	
	In summary, while prior works have demonstrated significant advancements across classical, CS-based, and DL-based channel estimation techniques, achieving an optimal trade-off between accuracy, overhead, and complexity remains an open challenge, particularly in wideband scenarios. These limitations motivate the development of our proposed deep learning framework, which aims to address these issues holistically.
			
	\subsection{Contributions of the Paper}
	While IRS-assisted channel estimation has attracted considerable attention, many existing works rely on oversimplified models and overlook critical challenges such as pilot overhead, computational complexity, limited generalization across varying channel conditions, and estimation accuracy. To bridge these gaps, we propose a novel multi-block attention (MBA) framework that jointly optimizes estimation accuracy, training efficiency, and generalization. Our design is specifically tailored to the structural properties of cascaded IRS channels in wideband OFDM systems. This ensures that both slot-dependent and subcarrier-dependent variations are thoroughly captured. The main contributions of this work are summarized as follows:
	
	\begin{itemize}				
		\item 
		\textbf{Pilot Overhead Reduction via Intelligent IRS Element Deactivation:} 
		We propose a novel strategy to reduce training overhead in IRS-assisted mmWave MIMO-OFDM systems by selectively deactivating a subset of IRS phase shifters (PSs). While deactivation may disrupt spatial correlation and weaken cascade channel observability \cite{10097678}, we address these effects using a DL–based reconstruction framework that recovers lost features. This approach is scalable and reduces pilot costs without compromising estimation accuracy.
		
		\item 
		\textbf{Multi-Block Axial Attention-Based Deep Learning Architecture:} We design an innovative two-stage deep neural network named MBA that integrates (i) a convolutional attention network (CAN) to reconstruct disrupted spatial correlations and (ii) a complex multi-convolutional network (CMN) for denoising. Unlike conventional architectures, MBA explicitly targets error mitigation and feature recovery from partially observed data. Moreover, our complexity analysis shows that the model scales linearly with IRS size, providing a scalable and efficient alternative to computationally expensive CS-based methods. Simulation results further validate the MBA’s superior performance and strong generalization capability across varying channel distributions. To the best of our knowledge, this is the first IRS channel estimation framework that leverages axial attention to reconstruct deactivated elements, combined with a two-stage recovery-denoising design to suppress error propagation directly. This approach is crucial for efficiency as it avoids the quadratic computational cost of applying standard self-attention to the entire flattened $N_t \times M$ channel matrix, a cost that would be prohibitive for large IRS arrays.
		
		\item 
		\textbf{Optimal Phase Shift Design for LS Estimation:} We derive the optimal IRS phase shift configuration for LS estimation in mmWave MIMO systems, demonstrating that DFT and Hadamard matrices minimize channel estimation's mean squared error (MSE). We analytically show that the LS estimator's MSE increases linearly with the IRS size, which is extremely inefficient in large-scale IRS deployments.
		
		\item 
		\textbf{Error Propagation Mitigation in Two-Stage DNNs:} 
		We conduct a theoretical analysis of error propagation in two-stage deep learning models, specifically within the context of IRS-assisted channel estimation \cite{bishop2006pattern}. We show how estimation errors accumulate and derive conditions under which our MBA model effectively suppresses this propagation. Both analysis and simulations confirm that MBA achieves significantly lower MSE than LS and state-of-the-art DL baselines.
	\end{itemize}
	
	\subsection{Organization and Notation}
	The remainder of this paper is organized as follows. Section \ref{sysModel} introduces the system model. In Section \ref{method}, we explain the proposed two-stage DNN architecture. Section \ref{theory} derives the optimal phase configuration for IRS cascade channel estimation. This is followed by theoretical analysis and complexity analysis of the proposed method. Simulation results are presented in Section \ref{sim}. Finally, Section \ref{conc} concludes the paper.
	
	\textit{Notation}: We employ boldface uppercase letters ($\textbf{A}$) to denote matrices, boldface lowercase letters ($\textbf{a}$) for vectors, and lowercase letters ($a$) for scalars. The symbols $(.)^T$, $(.)^H$, and $(.)^{-1}$ signify the transpose, Hermitian, and inverse of matrices, respectively. The notation $ \| \textbf{A} \|_F$ and $tr \{\textbf{A}\}$ represent the Frobenius norm and trace of matrix $\textbf{A}$, respectively.  Furthermore, $\mathbb{C}^{M \times N}$ denotes the complex space of dimension $M \times N$. Specifically, $\textit{diag}(\textbf{a})$ represents a diagonal matrix whose diagonal elements are the components of vector $\textbf{a}$. Lastly, $\mathbb{E}\left[.\right]$, shows the statistical expectation operator.
	
	\section{System Model}\label{sysModel}
	We consider an IRS-assisted MIMO OFDM system operating in time-division duplex (TDD) mode. We assume that each transmission comprises $T$ time slots, where each time slot corresponds to one OFDM symbol (block). The total duration of these time slots is shorter than the channel coherence time, so the channel is assumed to be quasi-static. Furthermore, due to the severe path loss and high susceptibility to blockages in mmWave communications, we assume that the direct link between the BS and MU is completely obstructed, rendering IRS-assisted transmission essential for reliable connectivity.
	\begin{figure}[t]
		\renewcommand{\figurename}{Figure}
		\centering
		\includegraphics[width=2.6in]{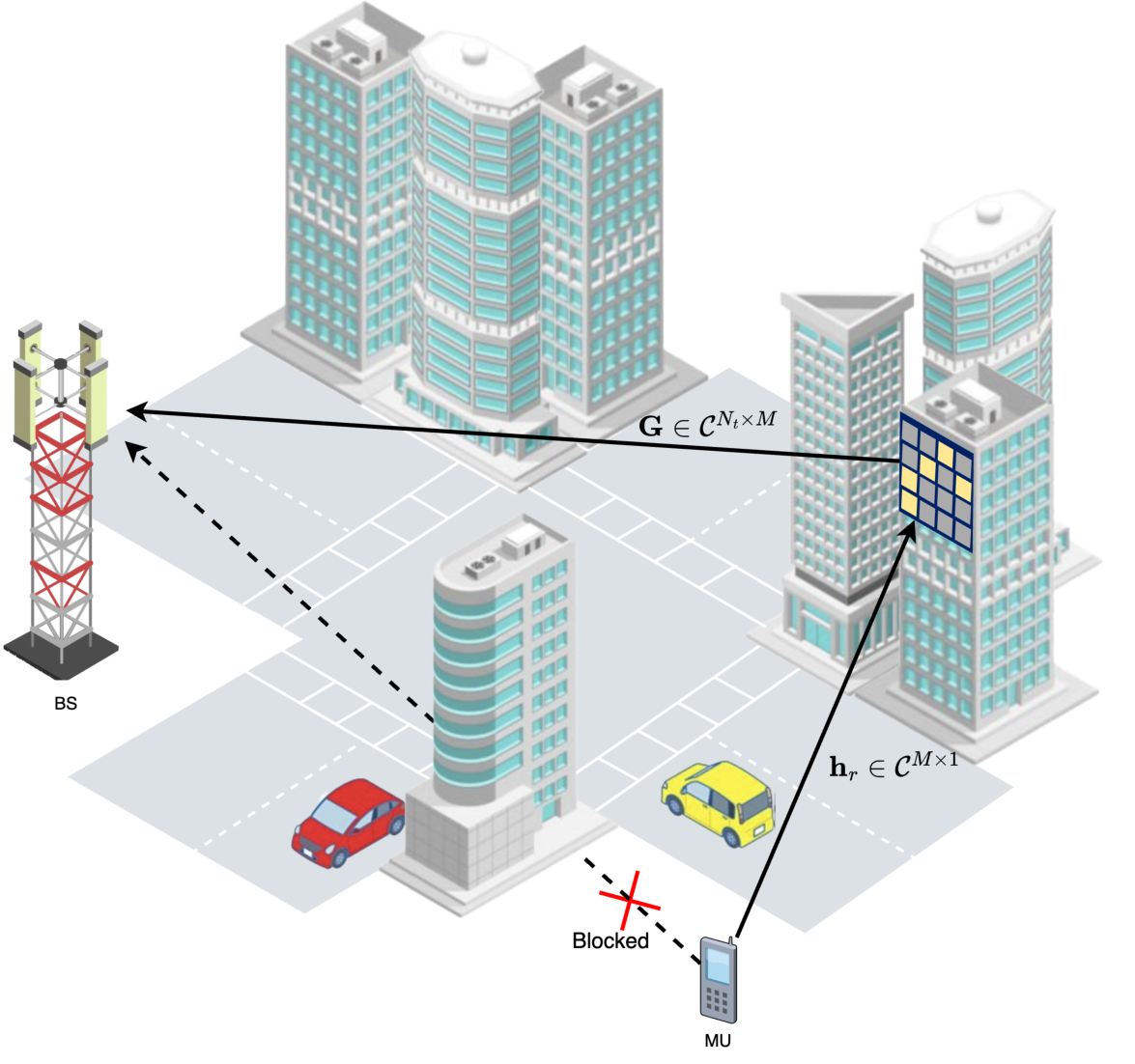}
		\caption{Schematic of IRS-assisted communication systems, where the direct channel between the BS and the MU is obstructed.}
		\label{fig:sys}
	\end{figure}
	
	As presented in Fig. \ref{fig:sys}, the BS is equipped with \(N_t\) antennas, while each user, with a single antenna, communicates with the BS through an IRS consisting of $M$ passive elements. Using orthogonal pilot signals, the BS estimates user channels. Without loss of generality, we focus on single-user channel estimation for simplicity.
	
	The communication channel between the IRS and the BS is represented by $\textbf{G} \in \mathbb{C}^{N_t \times M}$, and the channel between the MU and the IRS is denoted by $\textbf{h}_r \in \mathbb{C}^{M \times 1}$. The IRS reflection matrix at the $t^{\text{th}}$ time slot is defined as $\boldsymbol{\Phi}^{(t)} = \textit{diag}(\textbf{v}^{(t)}) \in \mathbb{C}^{M \times M}$, where $\textbf{v}^{(t)} \in \mathbb{C}^{M \times 1}$ is the IRS phase-shift vector:
	\begin{equation}\label{irs_vec}
		\textbf{v}^{(t)} = \big[\beta_1^{(t)}e^{j\theta_1^{(t)}}, \beta_2^{(t)}e^{j\theta_2^{(t)}}, \dots , \beta_M^{(t)} e^{j\theta_M^{(t)}} \big]^T.
	\end{equation}

	Here, the $i^{\text{th}}$ IRS element at the $t^{\text{th}}$ time slot is characterized by its phase shift $\theta_i^{(t)} \in [0, 2\pi)$ and amplitude reflection coefficient $\beta_i^{(t)} \in [0, 1]$. Specifically, $\beta_i^{(t)} = 0$ indicates that the element is deactivated, while $\beta_i^{(t)} = 1$ corresponds to full reflection \cite{8811733}. Each IRS element independently controls its phase shift and amplitude.
	
	\subsection{Channel Model}
	The channel considered in this paper is a multipath channel whose frequency response on each OFDM subcarrier needs to be estimated. To emulate practical propagation conditions, we adopt the clustered delay line (CDL) channel model specified in the 3GPP TR 38.901 standard based on extensive real-world measurement campaigns conducted in various environments and frequencies. The CDL model effectively captures multipath clustering, angular spreads, and delay dispersion, which are the dominant propagation characteristics in IRS-assisted OFDM systems. Following the guidelines in \cite{etsi_tr_138901_2022} and the modeling in \cite{8752012, 10309967}, the IRS-assisted channel is described in the delay domain as follows:
	\begin{equation}
		\begin{aligned}
			\label{ch_mod}
			\textbf{G}(\tau) = \sqrt{\frac{N_tM}{L_{\text{BS-IRS}}}}  \sum_{i=1}^{L_{\text{BS-IRS}}} \alpha_i \delta\left(\tau - \tau_i\right)\textbf{a}_{\text{BS}}(\phi^\text{BS}_{i})\\
			\times \textbf{a}^H_{\text{IRS}}(\phi^\text{IRS}_{i},\theta^\text{IRS}_{i}),
		\end{aligned}
	\end{equation}
	\begin{equation}\label{ch_mod2}
		\begin{aligned}
			\textbf{h}_r(\tau) = \sqrt{\frac{M}{L_{\text{IRS-MU}}}}  \sum_{j=1}^{L_{\text{IRS-MU}}} \alpha_j \delta(\tau - \tau_j)\textbf{a}_{\text{IRS}}(\phi^\text{IRS}_{j},\theta^\text{IRS}_{j}),
		\end{aligned}
	\end{equation}
	
	\noindent where $L_{\text{BS-IRS}}$ and $L_{\text{MU-IRS}}$ present the number of paths between the BS and IRS and between the MU and IRS, respectively. The delay and average power gain of \(i^\text{th}\) path are denoted by \(\tau_i\) and \(\alpha_i\), respectively where \(\alpha_i \sim \mathcal{CN}(0,\sigma_{\alpha}^2)\). The angles $\phi^{\text{IRS}}_{i}$ and $\theta^{\text{IRS}}_{i}$ represent the azimuth and elevation at the IRS. Similarly, $\phi^\text{BS}_{i}$ denotes the azimuth AoA or AoD at the BS. Furthermore, the IRS and BS array response vectors are denoted by $\textbf{a}_{\text{IRS}}$ and $\textbf{a}_{\text{BS}}$, respectively. In particular, the array response vector of a uniform linear array (ULA) with $N_{\text{ULA}}$ half-wavelength spaced elements is given by:
	\begin{equation}
		\begin{aligned}
			\label{eq:ULA}
			\textbf{a}_{\text{ULA}}(\phi)=\frac{1}{\sqrt{N_{\text{ULA}}}}\bigg[1,e^{j\pi\sin(\phi)},...,e^{j\pi(N-1)d\sin(\phi)}\bigg]^T,
		\end{aligned}
	\end{equation}
	\noindent and the array response vector of a half-wavelength spaced uniform planar array (UPA) with $N_{\text{UPA}} = N_x \times N_y$ elements is given by:
	\begin{equation}
		\label{eq:UPA}
		\begin{aligned}
			\textbf{a}_{\text{UPA}}(\phi,\theta)=\frac{1}{\sqrt{N_{\text{UPA}}}}\bigg[1,\ldots,e^{j\pi(n_x\sin(\phi)\sin(\theta)+n_y\cos(\theta)},\\\ldots,
			e^{j\pi((N_x-1)\sin(\phi)\sin(\theta)+(N_y-1)\cos(\theta)}\bigg]^T,
		\end{aligned}
	\end{equation}
	\noindent where $n_x$ and $n_y$ index the horizontal and vertical antenna elements.
	
	Based on the channel models in \eqref{ch_mod} and \eqref{ch_mod2}, the channel frequency response of the $k^{\text{th}}$ subcarrier for the BS–IRS and IRS–user channels is given by:
	\begin{equation}
		\begin{aligned}
			\label{G_freq}
			\textbf{G}_k = \sqrt{\frac{N_tM}{L_{\text{BS-IRS}}}}  \sum_{i=1}^{L_{\text{BS-IRS}}}\alpha_i e^{-j2\pi\tau_if_s\frac{k}{K}}\textbf{a}_{\text{BS}}(\phi^\text{BS}_{i})
			\textbf{a}^H_{\text{IRS}}(\phi^\text{IRS}_{i},\theta^\text{IRS}_{i}),
		\end{aligned}
	\end{equation}	
	\begin{equation}\label{ch_mod2_freq}
		\begin{aligned}
			\textbf{h}_{r_k} = \sqrt{\frac{M}{L_{\text{IRS-MU}}}}  \sum_{j=1}^{L_{\text{IRS-MU}}} \alpha_j e^{-j2\pi\tau_jf_s\frac{k}{K}}\textbf{a}_{\text{IRS}}(\phi^\text{IRS}_{j},\theta^\text{IRS}_{j}),
		\end{aligned}
	\end{equation}	
	
	\noindent where $f_s$ denotes the sampling rate, and $K$ represents the total number of subcarriers.
	
	\subsection{Cascaded Channel Model}
	In the uplink training phase, the single-antenna user transmits a pilot signal $s_k^{(t)}$ at the $t^{\text{th}}$ time slot on the $k^{\text{th}}$ subcarrier. The received signal at the BS, $\textbf{y}_k^{(t)} \in \mathbb{C}^{N_t\times 1}$, is expressed as follow:
	\begin{equation}\label{recieve}
		\begin{aligned}
			\textbf{y}_k^{(t)} &= \textbf{G}_k\boldsymbol \Phi^{(t)}\textbf{h}_{r_k} s_k^{(t)} + \textbf{n}_k^{(t)} \\
			&= \textbf{G}_k \textit{diag}(\textbf{h}_{r_k})\textbf{v}^{(t)} s_k^{(t)} + \textbf{n}_k^{(t)} \\
			&\triangleq \textbf{H}_{\text{cs}_k} \textbf{v}^{(t)} s_k^{(t)} + \textbf{n}_k^{(t)} ,
		\end{aligned}
	\end{equation}
	\noindent where $\textbf{n}_k^{(t)} \in \mathbb{C}^{N_t \times 1}$ is the additive noise vector at the BS, whose entries are independent and identically distributed circularly symmetric complex Gaussian variables with zero mean and variance $\sigma_n^2$. The cascaded channel matrix $\textbf{H}_{\text{cs}_k} \triangleq \textbf{G}_k\textit{diag}(\textbf{h}_{r_k}) \in \mathbb{C}^{N_t\times M}$ captures the joint effect of the BS–IRS channel $\textbf{G}_k$ and the IRS–user channel $\textbf{h}_{r_k}$.
	
	Downlink channels are acquired through uplink channel estimation by exploiting channel reciprocity in TDD systems. To estimate $\textbf{H}_{\text{cs}_k}$, the system requires sufficient observations with varying IRS phase shift configurations. Specifically, during $B$ consecutive training slots, the IRS applies distinct phase-shift vectors $\textbf{v}^{(t)}$, while the MU transmits pilot symbols $s_k^{(t)}$. After collecting these $B$ measurements, the BS reconstructs the cascaded channel matrix. A summary of the notation is provided in Table \ref{tab:not}.
		
	\subsection{Conventional LS Estimator for Cascaded Channel}
	The received signal matrix over $B$ time slots in the $k^{\text{th}}$ subcarrier can be presented as:	
	\begin{equation}\label{LS1}
		\begin{aligned}
			\hat{\textbf{Y}}_k  = \textbf{H}_{\text{cs}_k} \boldsymbol{\Psi} + \hat{\textbf{N}}_k,
		\end{aligned}
	\end{equation}
	\noindent where $\boldsymbol{\Psi} \triangleq \big[\textbf{v}^{(1)}, \dots, \textbf{v}^{(B)}\big] \in \mathbb{C}^{M\times B}$ is the IRS phase matrix across $B$ time slots, $\hat{\textbf{N}}_k \triangleq \big[\hat{\textbf{n}}_k^{(1)}, \dots, \hat{\textbf{n}}_k^{(B)}\big] \in \mathbb{C}^{N_t\times B}$ is the equivalent noise matrix, and $\hat{\textbf{Y}}_k \triangleq \big[\hat{\textbf{y}}_k^{(1)}, \dots, \hat{\textbf{y}}_k^{(B)}\big] \in \mathbb{C}^{N_t\times B}$ denotes the matrix of received signals. Each column is given by:
	\begin{equation}
		\begin{aligned}
			\hat{\textbf{y}}_k^{(t)} =  \frac{1}{p_k^{(t)}}\textbf{y}_k^{(t)} (s_k^{(t)})^{*} =  \textbf{H}_{\text{cs}_k} \textbf{v}^{(t)} + \hat{\textbf{n}}_k^{(t)},
		\end{aligned}
	\end{equation}
	\noindent where $p_k^{(t)}$ denoting the pilot signal power.

	The training dataset encompasses data from all subcarriers, eliminating the need to retrain the MBA model for each individual sub-carrier. The model is trained once and subsequently applied across all sub-carriers. As a result, runtime and computational complexity increase linearly with the number of sub-carriers. Using the conventional LS estimator, the cascaded channel $\textbf{H}_{\text{cs}_k}$ is then obtained as follows:
	\begin{equation}\label{LS_est}
		\begin{aligned}
			\hat{\textbf{H}}_{\text{LS}_k} = \hat{\textbf{Y}}_k\times\boldsymbol{\Psi}^H\left(\boldsymbol{\Psi}\boldsymbol{\Psi}^H\right)^{-1}.
		\end{aligned}
	\end{equation}
	
	\begin{figure*}[t]
		\renewcommand{\figurename}{Figure}
		\centering
		\subfloat[][]{\includegraphics[width=1in, height=1in]{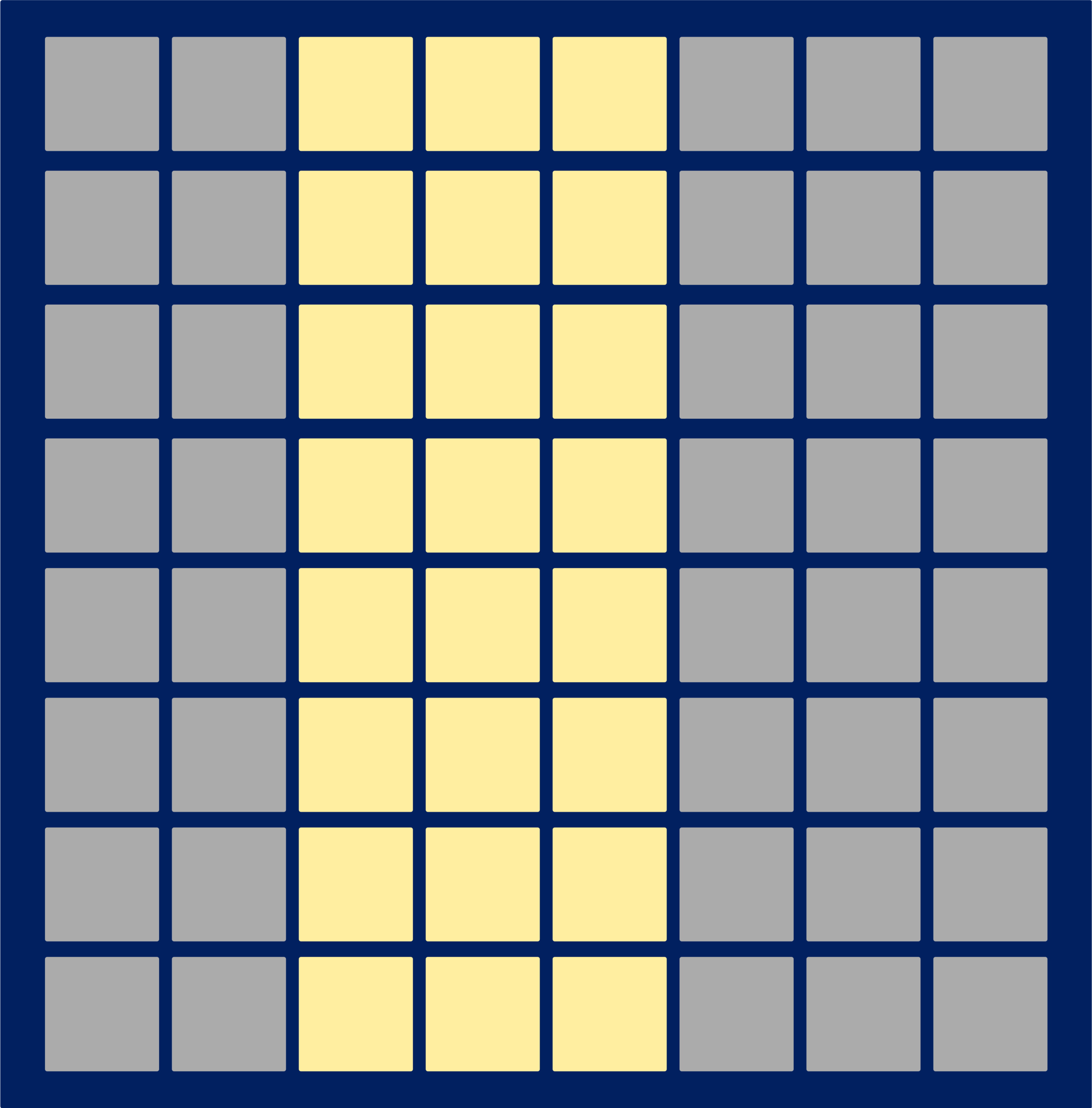}}\quad \quad\
		\subfloat[][]{\includegraphics[width=1in, height=1in]{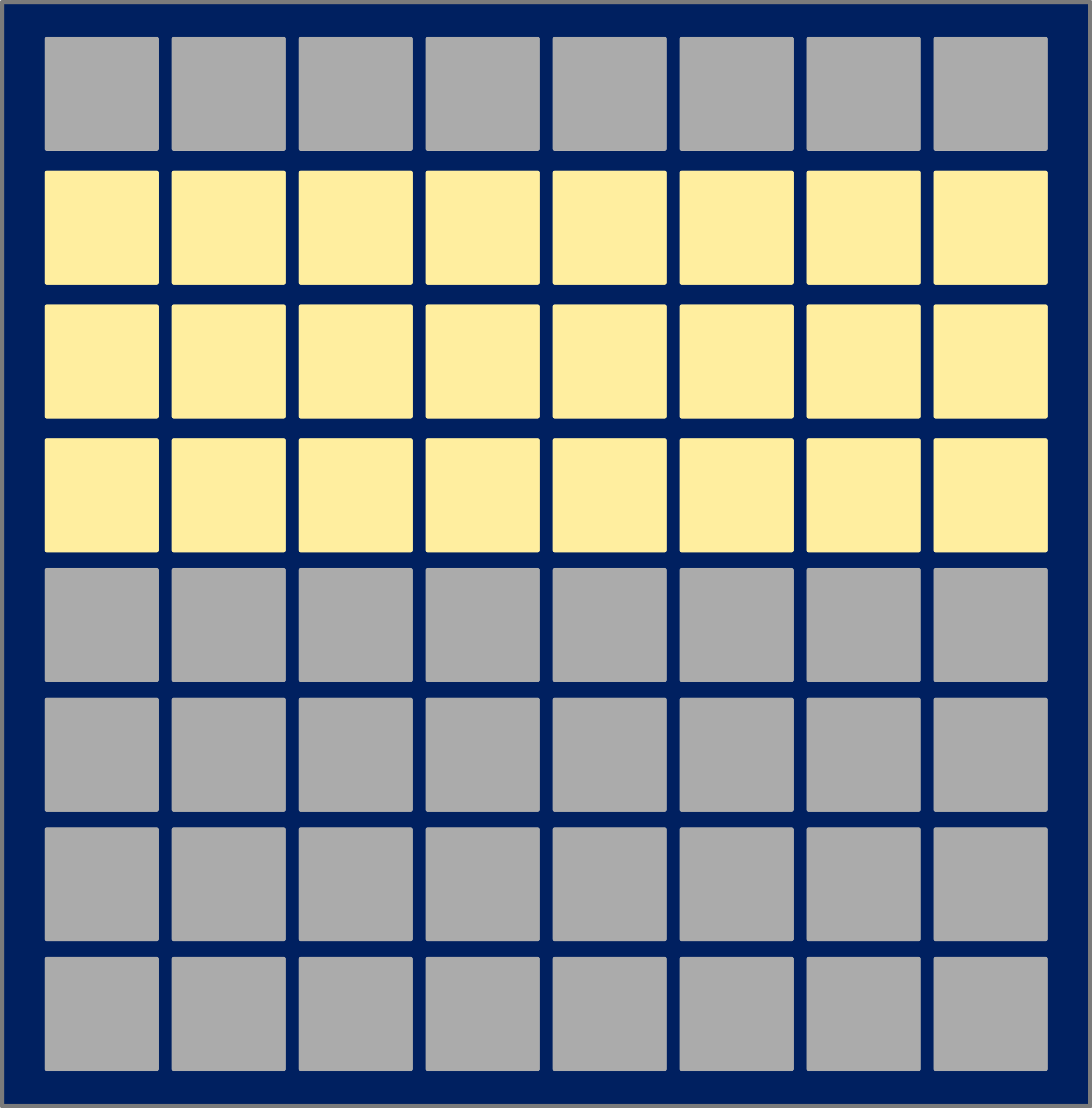}}\quad \quad\
		\subfloat[][]{\includegraphics[width=1in, height=1in]{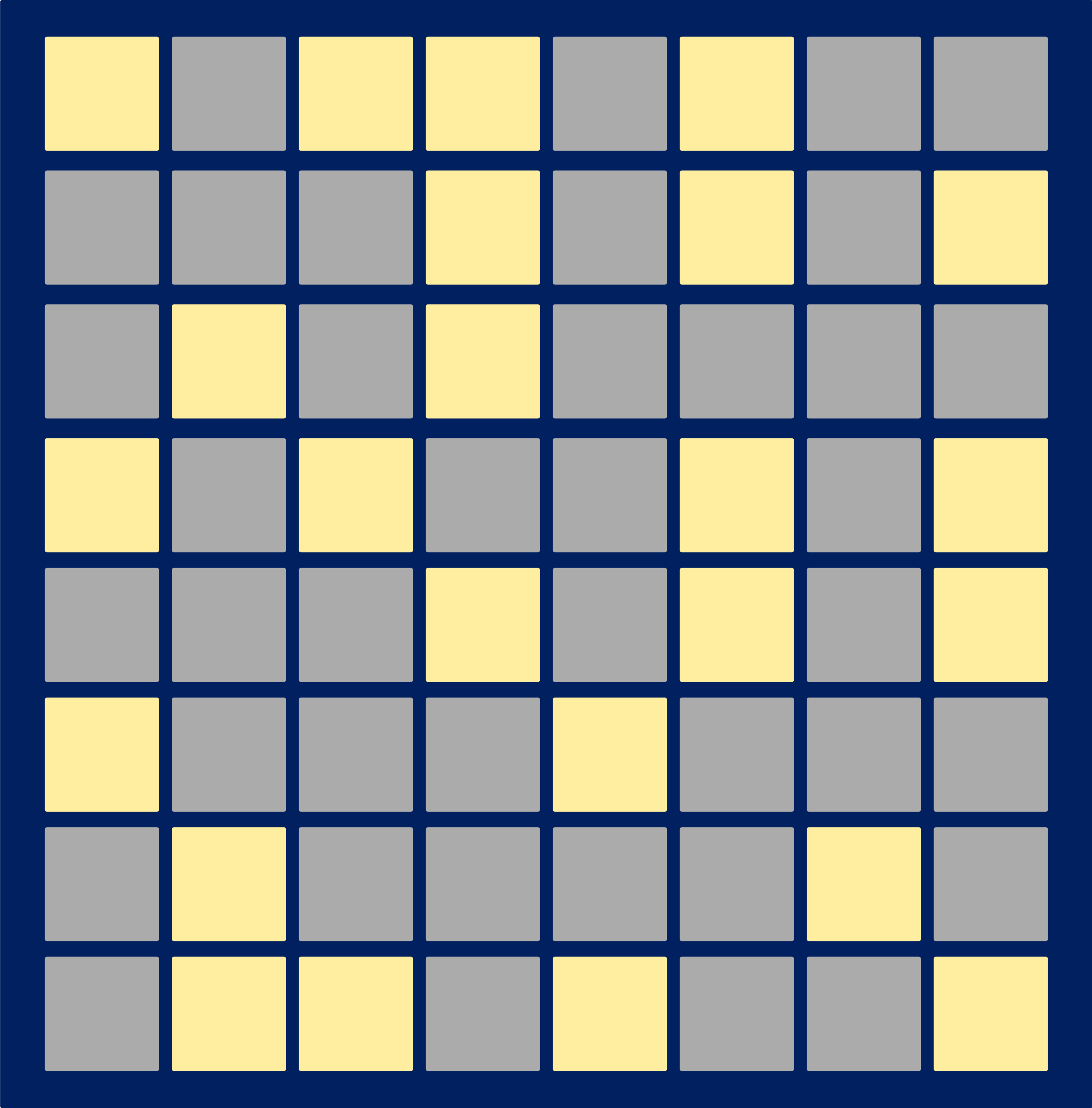}}\quad \quad\
		\subfloat[][]{\includegraphics[width=1in, height=1in]{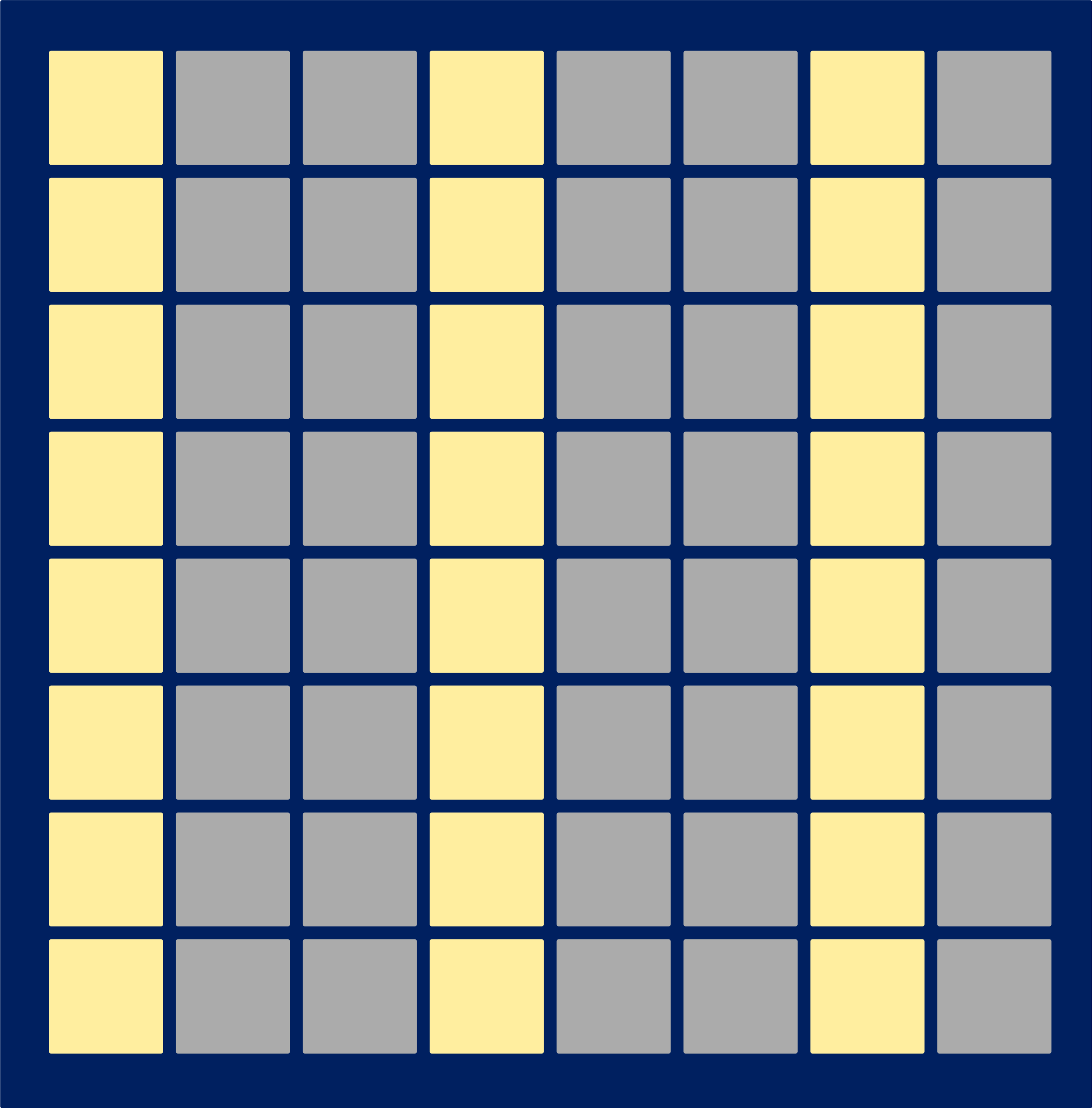}}
		\caption{Comparison of IRS activation patterns, where yellow represents activated elements, and gray represents deactivated ones. (a) Column-wise activation; (b) Row-wise activation; (c) Random scheme activation; (d) Proposed scheme.}
		\label{patterns}
	\end{figure*}  		
	
	\begin{table}
		\caption{System Model Notations}
		\label{tab:not}
		\begin{center}
			\scalebox{0.86}{
			\begin{tabular}{ c | c }
				\textbf{Notation} & \textbf{Definition} \\
				\noalign{\hrule height 1.5pt}
				$N_t$ &  Number of antennas at the BS. \\
				\hline
				$M$ & Number of elements in the IRS. \\ 
				\hline			 			
				$\textbf{G}_k\in \mathbb{C}^{N_t \times M}$ & BS-IRS channel matrix for the \(k^{\text{th}}\) subcarrier. \\
				\hline
				$\textbf{h}_{r_k}\in \mathbb{C}^{M \times 1}$ &IRS-MU channel vector for the $k^{\text{th}}$ subcarrier.\\
				\hline 		
				$\textbf{H}_{\text{cs}_k}\in \mathbb{C}^{N_t \times M}$ & Cascade channel matrix (BS-IRS-MU) for the $k^{\text{th}}$ subcarrier. \\
				\hline 				 			
				$\boldsymbol{\Phi}_t \in \mathbb{C}^{M\times M}$ & Diagonal reflection matrix of the IRS at time slot $t$ \\
				\hline 			
				$\textbf{v}_t\in \mathbb{C}^{M\times1}$ &IRS control vector at time slot $t$. \\
				\hline 		
				$\boldsymbol{\Psi}\in \mathbb{C}^{M\times B}$ &  IRS control vectors over $B$ time slots.\\
				\hline				
				$\hat{\textbf{Y}}_k\in \mathbb{C}^{N_t\times B}$ & Received signal matrix over \( B \) time slots in the \( k^{\text{th}} \) subcarrier.\\
				\hline	
				$\hat{\textbf{N}}_k\in \mathbb{C}^{N_t\times B}$ & The noise matrix over \( B \) time slots in the \( k^{\text{th}} \) subcarrier.\\
				\noalign{\hrule height 1.5pt}				
			\end{tabular}}
		\end{center}		
	\end{table}
	
	It is important to note that the use of the LS estimator, as defined in \eqref{LS_est}, is valid only under the condition $B\geq M$, where $B$ denotes the number of pilot symbols and $M$ represents the number of IRS elements. This constraint becomes particularly significant when the total number of available time slots $T$ (i.e., OFDM symbols) is smaller than $M$. In such cases, channel estimation using \eqref{LS_est} becomes impractical, as the channel may vary across different coherence intervals, violating the quasi-static assumption. Furthermore, the LS estimator is highly sensitive to noise, which can lead to substantial estimation errors. These limitations highlight the necessity of developing more efficient cascaded channel estimation techniques that can reduce training overhead while maintaining estimation accuracy.

	\section{Proposed Method}\label{method}
	This section presents the two-stage MBA framework, detailing its architecture, key components, and application to IRS-assisted channel estimation. 
	
	\subsection{Activation Pattern for IRS}
	A key challenge in IRS-assisted channel estimation is the high training overhead resulting from a large number of passive elements. To address this, we selectively deactivate certain IRS elements, effectively reducing the dimensionality of the cascaded channel and enabling more efficient channel estimation using an LS estimator.
	
	In this framework, the deactivation or activation of the $i^{\text{th}}$ IRS element in $t^\text{th}$ time slot is achieved by setting $\beta_i^{(t)} = 0$ or $\beta_i^{(t)} = 1$, respectively. Let $B$ denote the number of activated elements. Since $M-B$ columns of $\textbf{H}_{\text{cs}_k}$, which correspond to deactivated elements, do not contribute to the received signal, we can eliminate them and use the reduced dimension cascade channel between MU and BS denoted by $\tilde{\textbf{H}}_{\text{cs}_k}\in \mathbb{C}^{N_t \times B}$. The received signal over $B$ time slots in $k^\text{th}$ subcarrier is then given by:
	\begin{equation}\label{LS2}
		 \tilde{\textbf{Y}}_k  = \tilde{\textbf{H}}_{\text{cs}_k} \tilde{\boldsymbol{\Psi}} + \hat{\textbf{N}}_k, 
	\end{equation}
	
	\noindent where $\tilde{\boldsymbol{\Psi}} \in \mathbb{C}^{B \times B}$ represents the IRS matrix associated with the $B$ activated PS elements. Hence, the LS estimation $k^\text{th}$ $\tilde{\textbf{H}}_{\text{cs}_k}$ is obtained as:
	\begin{equation}\label{LS_est2} \hat{\tilde{\textbf{H}}}_{\text{LS}_k} = \tilde{\textbf{Y}}_k\tilde{\boldsymbol{\Psi}}^H\left(\tilde{\boldsymbol{\Psi}} \tilde{\boldsymbol{\Psi}}^H\right)^{-1}. \end{equation}
	
	Although deactivating IRS elements effectively reduces pilot overhead, it disrupts spatial correlations and introduces NLOS effects \cite{10097678}. Thus, we develop a deep learning-based recovery mechanism using the MBA framework to address this issue. 
	
	Figure \ref{patterns} illustrates four IRS activation patterns. The proposed scheme ensures that at least one fully active column is maintained while deactivated elements are distributed evenly, preserving local spatial correlation.
	
	\subsection{Dataset Construction}
	To construct the dataset, CSI samples are generated for each of the $K$ subcarriers using the channel model in Section \ref{sysModel}, with channel estimation performed independently per subcarrier. For subcarrier $k$, the reduced-dimension channel $\tilde{\textbf{H}}_{\text{cs}_k}$ is first obtained via the LS estimator in \eqref{LS_est2}. Zero columns are then inserted at the deactivated PS indices to recover the missing entries, yielding the augmented estimate $\hat{\textbf{H}}_{\text{Aug}_k}$.
	
	Since $\textbf{H}_{\text{cs}_k}$ is complex-valued, its real and imaginary parts are separated and concatenated into a real-valued tensor of size $(N_t, M, 2)$. Repeating this process over all $K$ subcarriers produces a dataset in which each sample corresponds to the channel of a single subcarrier. This dataset is subsequently used to train the MBA model to refine the noisy augmented estimate $\hat{\textbf{H}}_{\text{Aug}_k}$ toward the true channel $\textbf{H}_{\text{cs}_k}$.
	
	\subsection{Attention Mechanism}
	As stated earlier, to overcome the serious issue of huge overhead in traditional channel estimation approaches, we deactivate some IRS elements at the cost of disrupting the spatial correlation. To solve this problem, we employ an attention mechanism as part of our proposed channel estimation method for the following reasons:
	\begin{itemize}
		\item 
		\textbf{Spatial correlation compensation:} The attention mechanism can capture long-range spatial correlation by dynamically reweighting features \cite{9526282}. This can improve the estimation accuracy in dynamic environments \cite{9526282}.
		
		\item 
		\textbf{Adaptability:} The attention mechanism can model global dependencies. Thus, they can effectively handle scattered paths and non-stationary channels.
		
		\item 
		\textbf{Parallelism:} Another key advantage of these mechanisms, specifically self-attention, is parallelism. Unlike sequential processing, self-attention allows parallel computation, significantly accelerating training and inference.
	\end{itemize}
	
	The self-attention mechanism that we incorporate relies on three key components: Key (\textbf{K}), Query (\textbf{Q}), and Value (\textbf{V}). The input $\textbf{X} \in \mathbb{R}^{N_t \times M}$ is first transformed using three multilayer perceptrons (MLPs) to generate the corresponding weights in the original attention mechanism \cite{vaswani2017attention}. To reduce complexity, we replace traditional fully connected layers in self-attention with multi-convolutional blocks (MB), which better exploit spatial correlations in IRS-based channel estimation.
	
	Compared to MLPs, convolutional neural networks (CNNs) require less training data, exhibit stronger noise robustness, and are better suited for structured inputs \cite{rolnick2017deep}. While attention mechanisms are effective for modeling long-range dependencies, most IRS-based tasks primarily rely on local correlations, making CNNs a more practical choice. In our design, convolutional layers improve efficiency and reduce self-attention complexity.
	
	In the proposed model, to address the challenge of missing channel information resulting from the deactivation of IRS elements, we employ a specialized deep learning mechanism known as axial attention \cite{ho2019axial}. The primary goal of this mechanism is to intelligently reconstruct the deactivated elements by learning the spatial relationships and interdependencies from the active ones.

	The process begins with the initial, incomplete channel estimate, which is structured as a three-dimensional tensor, denoted as $\textbf{X}$. This tensor has dimensions of $N_t \times M \times 2$, where $N_t$ represents the number of antennas at BS, $M$ is the total number of reflecting elements on the IRS, and the final dimension of size 2 separates the real and imaginary parts of the complex channel values for processing by the neural network. Next, the model transforms the input tensor $\textbf{X}$ into three distinct tensors: \textbf{Q}, \textbf{K}, and \textbf{V}. These tensors are generated by projecting the input $\textbf{X}$ through learned weight matrices ($\textbf{W}_q, \textbf{W}_k, \textbf{W}_v$), as shown below:
	\begin{equation}
		\begin{aligned}
			\textbf{Q} = \textbf{X}\textbf{W}_q, \quad \textbf{K} = \textbf{X}\textbf{W}_k, \quad \textbf{V} = \textbf{X}\textbf{W}_v.
		\end{aligned}
	\end{equation}
	
	Each resulting tensor has dimensions $N_t \times M \times d$, where $d$ is a learned feature dimension. The core of the mechanism lies in calculating attention scores to determine the focus each IRS element should place on every other element. A relevance score is computed by comparing the Query of one element with the Key of all other elements via a dot product, $[\textbf{Q}\textbf{K}^T]_n = \textbf{Q}_n \textbf{K}_n^T$. Importantly, this calculation is carried out independently for each of the $N_t$ BS antennas. This approach concentrates the model's learning capacity along the IRS element dimension ($M$), precisely where information was lost. The resulting raw scores are scaled by $1/\sqrt{d}$ for numerical stability and then passed through a Softmax function. This function converts the scores into a set of weights that sum to one, yielding the final attention tensor $\textbf{S} = \text{Softmax}(\textbf{Q}\textbf{K}^T/\sqrt{d})$ with dimensions $N_t \times M \times M$. To formalize the IRS indexing, let $\underline{i} = (r, c)$ denote the $i^{\text{th}}$ IRS element located at the $r^{\text{th}}$ row and $c^{\text{th}}$ column of the 2D planar IRS. Accordingly, an entry $\textbf{S}_{n, \underline{i}, \underline{j}}$ quantifies the importance of the element at position $\underline{j} = (r', c')$ in reconstructing the element at position $\underline{i} = (r, c)$ for the channel link associated with the $n^\text{th}$ BS antenna.
	
	Finally, these attention weights are used to generate a refined and complete channel estimate. The attention tensor $\textbf{S}$ is applied to the Value tensor (\textbf{V}). This process enables a deactivated element, initially represented by zeros, to be reconstructed using an informed combination of features from the most relevant active elements. The final output, $\textbf{O}$, is obtained by $\textbf{O} = \textbf{S}\textbf{V}$. As a result, the attention mechanism is computed as follows:
	\begin{equation}
		\begin{aligned}
			\text{Attention}\left(\textbf{X}\right) = \text{Softmax}\left(\dfrac{\textbf{Q}  \textbf{K}^T}{\sqrt{d}}\right)\textbf{V}.
		\end{aligned}
	\end{equation}
	
\begin{figure}
		\renewcommand{\figurename}{Figure}
		\centering
		\includegraphics[width=3.5in, height=1.3in]{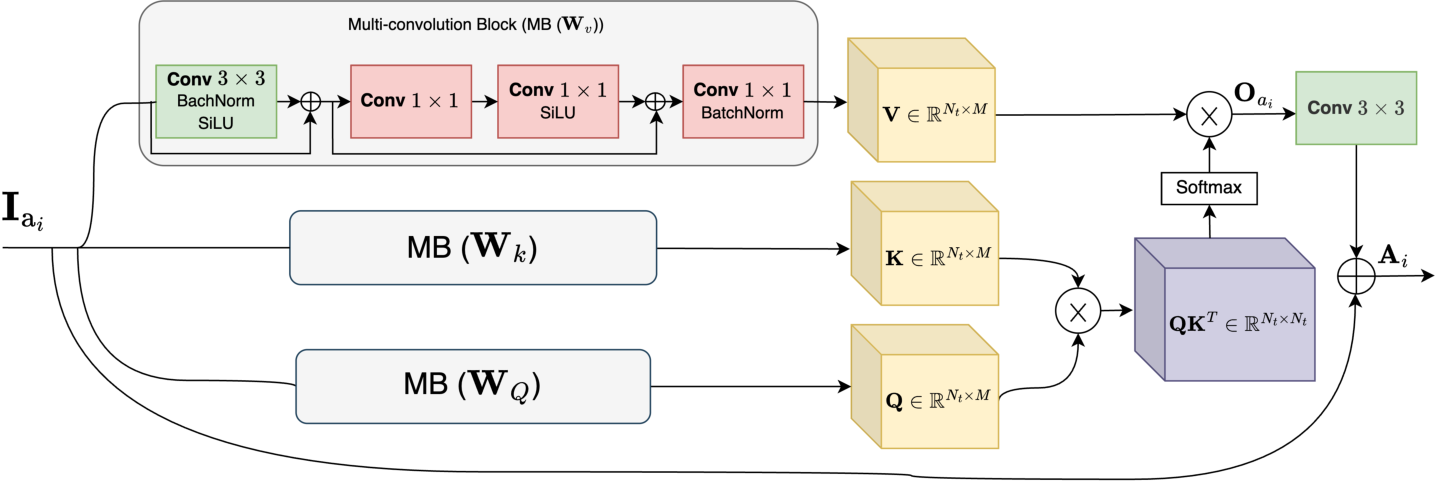}
		\caption{Structure of the Attention Block (AB), combining a self-attention mechanism and a multi-convolutional block (MB) to enhance feature extraction.}
		\label{fig:attention}
	\end{figure}

	The resulting tensor $\textbf{O}$ represents the reconstructed channel with the missing information filled in. This axial approach is significantly more computationally efficient than standard attention mechanisms, ensuring the model remains scalable for large-scale IRS deployments. Finally, the schematic representation of the attention mechanism is shown in Fig. \ref{fig:attention}.

	\subsection{Multi-Block Attention (MBA)}
\begin{figure}[t]
		\renewcommand{\figurename}{Figure}
		\centering
		\includegraphics[width=2.6in, height=3.5in]{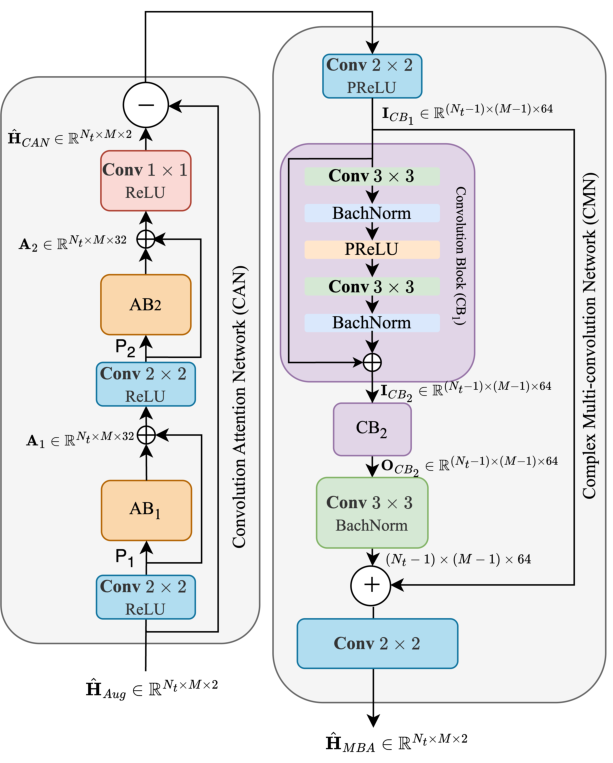}
		\caption{Overview of the proposed MBA framework.}
		\label{fig:MBA}
	\end{figure}

	As presented in Fig. \ref{fig:MBA}, the proposed MBA framework comprises two primary components: (i) the CAN for restoring spatial correlations and (ii) the CMN for mitigating noise. The CAN model is initially trained in isolation, after which its parameters are frozen before training the CMN module.
	
	The CAN architecture comprises two attention blocks (ABs) reconstructing inactive channels by capturing the missing spatial correlations. Each attention block $\textbf{A}_i$ ($i = 1,2$) is composed of a convolutional layer, a ReLU activation, and a self-attention mechanism, as formulated below:
	\begin{equation}
		\begin{aligned}
			\textbf{A}_i &= \textbf{I}_{\text{a}_i} + \text{Conv}(\text{Attention}(\textbf{I}_{\text{a}_i})),
		\end{aligned}
	\end{equation}
	\noindent where $\textbf{I}_{\text{a}_i}$ denotes the input to the attention mechanism. The output representation from the CAN module is mathematically expressed as follows:
	
	\begin{equation}
		\begin{aligned}
			\mathbf{P}_1 &= \text{ReLU}\big(\text{Conv}(\hat{\mathbf{H}}_{\text{Aug}_k})\big), \\
			\textbf{A}_1 &= \text{AB}_1(\textbf{P}_1) + \textbf{P}_1,\\
			\mathbf{P}_2 &= \text{ReLU}\big(\text{Conv}(\mathbf{A}_1)\big), \\
			\textbf{A}_2 &= \text{AB}_2(\textbf{P}_2) + \textbf{P}_2,\\
			\hat{\textbf{H}}_{\text{CAN}_k}  &=\hat{\mathbf{H}}_{\text{Aug}_k} - \text{ReLU}(\text{Conv}(\textbf{A}_2)),
		\end{aligned}
		\label{eq:can_module}
	\end{equation}
	
	\noindent where $\hat{\textbf{H}}_{\text{Aug}_k}$ refers to the channel matrix with augmented zero columns described in Section~\ref{method}.B, and $\textbf{P}_1$, $\textbf{P}_2$ serve as the inputs to $\text{AB}_1$ and $\text{AB}_2$, respectively.
	
	The second component, CMN, suppresses noise and residual distortions in the channel estimates produced by CAN. It is designed with a deep-layered architecture incorporating residual connections to counteract vanishing gradient issues during training. The CMN module consists of two convolutional blocks $\text{CB}_i$ ($i = 1,2$), each defined as:
	\begin{equation}
		\begin{aligned}
			\text{CB}_i (\textbf{I}_i) = \text{BN}(\text{Conv}(\text{PReLU}(\text{BN}(\text{Conv}(\textbf{I}_i))))) + \textbf{I}_i,
		\end{aligned}
	\end{equation}
	\noindent where $\text{BN}$ denotes batch normalization, $\textbf{I}_i$ is the input, and the output of each CB is added to its input as a residual connection.
	
	The full output of the CMN module is computed as:
	\begin{equation}
		\begin{aligned}
			\mathbf{O}_{\text{CMN}} &= \text{Conv}\big(\text{BN}(\text{Conv}(\text{CB}_2(\text{CB}_1( \\
			&\quad \text{PReLU}(\text{Conv}(\hat{\mathbf{H}}_{\text{CAN}_k})) ))))\big) + \text{Conv}(\hat{\mathbf{H}}_{\text{CAN}_k}),
		\end{aligned}
		\label{eq:cmn_module}
	\end{equation}
	\noindent where $\textbf{O}_{\text{CMN}}$ denotes the refined channel estimation output.
	
	The final output of the MBA model is obtained by sequentially applying CAN and CMN, ensuring robust reconstruction and denoising. This process is described as:
	\begin{equation}
		\hat{\mathbf{H}}_{\text{MBA}_k} = \mathbf{O}_{\text{CMN}} = \mathcal{F}_{\text{CMN}}\big(\mathcal{F}_{\text{CAN}}(\hat{\mathbf{H}}_{\text{Aug}_k})\big),
		\label{eq:mba_output}
	\end{equation}
	
	\noindent where $\mathcal{F}_{\text{CAN}}(\cdot)$ and $\mathcal{F}_{\text{CMN}}(\cdot)$ represent the learned transformations of the CAN and CMN networks, respectively. This hierarchical design guarantees structural consistency and robustness to noise in the estimated channels.
	
	To optimize both networks, we adopt the MSE loss. The loss functions for CAN and CMN are defined as:
	\begin{equation}\label{loss_can}
		\begin{aligned}
			\mathcal{L}_{\text{CAN}} = \frac{1}{2D}\sum_{i=1}^{D} \left\| \textbf{H}_{\text{cs}_k}^{(i)} - \hat{\textbf{H}}_{\text{CAN}_k}^{(i)} \right\|_F^2,
		\end{aligned}
	\end{equation}
	\begin{equation}\label{loss_cmn}
		\begin{aligned}
			\mathcal{L}_{\text{CMN}} = \frac{1}{2D}\sum_{i=1}^{D} \left\| \textbf{H}_{\text{cs}_k}^{(i)} - \textbf{O}_{\text{CMN}_k}^{(i)} \right\|_F^2,
		\end{aligned}
	\end{equation}
	
	\noindent where $D$ is the total number of training samples.
	
	\section{Theoretical Analysis}\label{theory}
	This section presents two key theoretical contributions to IRS-assisted channel estimation. First, we derive the optimal phase for the IRS elements for the channel estimation. Second, we analyze the error propagation phenomenon in two-stage DNNs and theoretically demonstrate how our proposed architecture effectively suppresses this detrimental effect.
	\subsection{On the optimality of IRS matrix}  
	As previously mentioned, it is essential to determine the phase of the PS in the IRS for each time slot during the estimation stage. The phase of the PS in the LS estimator, as indicated in \eqref{LS_est}, plays a crucial role in the accuracy of channel estimation. Therefore, optimizing the IRS phase shifts in each time slot is vital for minimizing the MSE in the following optimization problem:
	\begin{equation*}
		\begin{aligned}
			\min_{\boldsymbol{\Psi}} \quad & \mathcal{J}_{\text{LS}} = \mathbb{E}\left[||\textbf{H}_{\text{cs}} - \hat{\textbf{H}}^{\text{LS}}_{\text{cs}} ||^2_F\right]\\
			\textrm{s.t.} \quad & |\boldsymbol{\Psi}(i,j)|=1, \quad i,j=1,2,...,M.
		\end{aligned}
	\end{equation*}
	
	By substituting \eqref{LS_est} into the objective function, we obtain:
	\begin{equation}\label{ls1}
		\begin{aligned}
			\mathcal{J}_{\text{LS}} &= \mathbb{E}\left[||\textbf{H}_{\text{cs}} - \left(\textbf{H}_{\text{cs}} \boldsymbol{\Psi} + \textbf{N}\right)\boldsymbol{\Psi}^{\dag} ||^2_F\right]\\
			&=\mathbb{E}\left[||\textbf{N}\boldsymbol{\Psi}^{\dag}||^2_F\right] \\&= tr \left\{       (\boldsymbol{\Psi}\boldsymbol{\Psi}^H)^{-1}\boldsymbol{\Psi} \mathbb{E} \left[\textbf{N}^H\textbf{N}\right] \boldsymbol{\Psi}^H(\boldsymbol{\Psi}\boldsymbol{\Psi}^H)^{-1}     \right\}.
		\end{aligned}
	\end{equation}
	
	Since each of the \(N_t\) rows of \(\textbf{N}\) has a covariance of \(\sigma_n^2 \textbf{I}_M\), we have \(\mathbb{E}[\textbf{N}^H \textbf{N}] = N_t \sigma_n^2 \textbf{I}_M\). Therefore, in equation \eqref{ls1}, by substituting this result, we obtain:
	\begin{equation}\label{ls2}
		\begin{aligned}
			\mathcal{J}_{\text{LS}} &= N_t \sigma_n^2 tr \left\{       (\boldsymbol{\Psi}\boldsymbol{\Psi}^H)^{-1}\boldsymbol{\Psi} \textbf{I}_M \boldsymbol{\Psi}^H(\boldsymbol{\Psi}\boldsymbol{\Psi}^H)^{-1}     \right\}\\ 
			&= N_t\sigma_n^2 tr\left\{ (\boldsymbol{\Psi}\boldsymbol{\Psi}^{H})^{-1} \right\}.
		\end{aligned}
	\end{equation}
	
	The optimization problem is now reformulated as follows:
	\begin{equation*}
		\begin{aligned}
			\min_{\boldsymbol{\Psi}} \quad & tr\left\{ (\boldsymbol{\Psi}\boldsymbol{\Psi}^{H})^{-1} \right\}\\
			\textrm{s.t.} \quad & |\boldsymbol{\Psi}(i,j)|=1, \quad i,j=1,2,...,M.
		\end{aligned}
	\end{equation*}
	
	Minimizing the trace of the inverse requires the eigenvalues of $\boldsymbol{\Psi}\boldsymbol{\Psi}^H$ to be uniformly distributed. The unimodular constraint requires $|\boldsymbol{\Psi}(i,j)| = 1$, implying $\boldsymbol{\Psi}(i,j) = e^{j\theta_{\textit{ij}}}$, with elements on the unit circle \cite{dujardin2017dynamical}. The optimal $\boldsymbol{\Psi}$ leverages equiangular tight frame (ETF) properties, which minimize the trace of the inverse of Gram matrices under unit-norm constraints. Thus, minimizing $tr\left\{(\boldsymbol{\Psi}\boldsymbol{\Psi}^{H})^{-1} \right\}$ for a unimodular $\boldsymbol{\Psi}$ involves constructing $\boldsymbol{\Psi}$ as an ETF, ensuring uniform eigenvalues of $\boldsymbol{\Psi}\boldsymbol{\Psi}^{H}$. Therefore, we choose $\boldsymbol{\Psi}$ as follows:
	\begin{equation}
		\begin{aligned}
			\boldsymbol{\Psi}\boldsymbol{\Psi}^H = \textbf{I}_M.
		\end{aligned}
	\end{equation}
	
	Two well-known examples of an ETF are the DFT and the Hadamard matrix. The Hadamard matrix is particularly advantageous when the PSs of the IRS are quantized, as it maintains a structured and unimodular design. When the DFT matrix is selected as the $\boldsymbol{\Psi}$ matrix, the MSE for the LS estimator can be expressed as follows:
	\begin{equation}
		\begin{aligned}
			\mathcal{J}_{\text{LS}} = MN_t\sigma_n^2.
		\end{aligned}
	\end{equation}	
	
	As we can see, the LS estimator's MSE scales linearly with IRS size due to the accumulation of noise across the elements.
	
	\subsection{Error propagation analysis}
	The conventional LS estimator provides a simple initial channel estimate but is highly sensitive to noise, particularly when IRS elements are deactivated during estimation. This leads to substantial estimation error, denoted as $\mathcal{J}_{\text{LS}}$. To address this issue, we introduce the MBA framework, a two-stage DL approach that enhances estimation robustness more effectively than single-stage models \cite{bishop2006pattern}. The first stage, the CAN, refines the LS estimate using self-attention to recover lost spatial features. Its output is modeled as follows:
	\begin{equation}\label{LS_CAN}
		\begin{aligned}
		\hat{\mathbf{H}}_{\text{CAN}} = \hat{\mathbf{H}}_{\text{LS}} + \lambda_{\text{CAN}} (\mathbf{H}_{\text{cs}} - \hat{\mathbf{H}}_{\text{LS}}) + \boldsymbol{\varepsilon}_{\text{CAN}},
		\end{aligned}
	\end{equation}
	
	\noindent where $\lambda_{\text{CAN}}$ is the feature recovery gain, and $\boldsymbol{\varepsilon}_{\text{CAN}}$ denotes the residual error due to imperfect recovery. This structure is analogous to classical error correction models, where the network learns to reconstruct missing information based on prior estimates.
	
	The term $\mathbf{H}_{\text{cs}} - \hat{\mathbf{H}}_{\text{LS}}$ represents the estimation error, guiding CAN to refine rather than relearn the channel. If $\lambda_{\text{CAN}} = 1$, CAN ideally recovers the complete channel, while $\lambda_{\text{CAN}} \leq 0$ implies no improvement or degradation. By substituting the LS estimation model into \eqref{LS_CAN}, the residual error after CAN processing is given by
	\begin{equation}\label{LS_CAN2}
		\boldsymbol{\varepsilon}_{\text{CAN}} = (1 - \lambda_{\text{CAN}}) \hat{\mathbf{N}}.
	\end{equation}
	
	In this study, the NMSE serves as the performance metric; therefore, the feature recovery gain of CAN is defined as:
	\begin{equation}\label{LAMDA_attn}
		\lambda_{\text{CAN}} = \frac{\text{NMSE}_{\text{LS}} - \text{NMSE}_{\text{CAN}}}{\text{NMSE}_{\text{LS}}},
	\end{equation}	
	
	\noindent where the numerator quantifies the reduction in estimation error, and the normalization ensures scale invariance across different configurations. This approach aligns with the denoising autoencoder framework proposed in \cite{Vincent2018}, which measures robustness improvement via reductions in reconstruction error.

	Following CAN processing, the refined estimate $\hat{\mathbf{H}}_{\text{CAN}}$ is fed into the CMN module, which further suppresses residual noise using deep convolutional filtering. The output of CMN is modeled as follows:
	\begin{equation}\label{LS_CMN}
		\hat{\mathbf{H}}_{\text{CMN}} = \hat{\mathbf{H}}_{\text{CAN}} + \lambda_{\text{CMN}} (\mathbf{H}_{\text{cs}} - \hat{\mathbf{H}}_{\text{CAN}}) + \boldsymbol{\varepsilon}_{\text{CMN}},
	\end{equation}
	
	\noindent where $\lambda_{\text{CMN}}$ captures the noise suppression gain, and $\boldsymbol{\varepsilon}_{\text{CMN}}$ is the residual error. Similar to CAN, the residual error after CMN can be expressed as:
	\begin{equation}\label{LS_MBA}
		\boldsymbol{\varepsilon}_{\text{CMN}} = (1 - \lambda_{\text{CMN}}) \boldsymbol{\varepsilon}_{\text{CAN}} = (1 - \lambda_{\text{CMN}})(1 - \lambda_{\text{CAN}}) \hat{\mathbf{N}}.
	\end{equation}
	
The noise suppression gain $\lambda_{\text{CMN}}$ is defined analogously to \eqref{LAMDA_attn} as follow:
	\begin{equation}\label{LAMDA_CMN}
		\lambda_{\text{CMN}} = \frac{\text{NMSE}_{\text{CAN}} - \text{NMSE}_{\text{CMN}}}{\text{NMSE}_{\text{CAN}}}.
	\end{equation}
	
	This expression follows the noise reduction principles observed in deep convolutional denoising networks \cite{zhang2017beyond}. The overall estimation error after both CAN and CMN stages can thus be characterized as follows:
	\begin{equation}\label{J_MBA}		
	\begin{aligned}
		\mathcal{J}_{\text{MBA}} &= \mathbb{E} \left[ \| \mathbf{H}_{\text{cs}}  - \hat{\mathbf{H}}_{\text{CMN}} \|_F^2 \right]\\
		&= \mathbb{E} \left[\|(1 - \lambda_{\text{CMN}})(1 - \lambda_{\text{CAN}}) \hat{\mathbf{N}}\|_F^2\right]\\
		 &=(1 - \lambda_{\text{CMN}})^2(1 - \lambda_{\text{CAN}})^2\mathbb{E} \left[\|\hat{\mathbf{N}}\|_F^2\right]\\
		&=(1-\lambda_{\text{CAN}})^2 (1 - \lambda_{\text{CMN}})^2 \mathcal{J}_{\text{LS}}.
	\end{aligned}	
	\end{equation}
	
	This formulation shows that as $\lambda_{\text{CAN}}, \lambda_{\text{CMN}} \to 1$, the error $\mathcal{J}_{\text{MBA}} \to 0$, indicating near-optimal channel reconstruction. Simulations validate this trend, demonstrating the superior performance of MBA over LS in terms of estimation accuracy. Moreover, increased pilot signal availability drives $\lambda_{\text{CAN}}$ and $\lambda_{\text{CMN}}$ toward unity.
	
	\subsection{Computational Complexity Analysis}
	Evaluating computational complexity is critical for assessing the real-time feasibility of IRS-based channel estimation techniques. This issue is crucial in wireless systems, where applications demand extremely low latencies. Here, we analyze the computational complexity of our proposed MBA-based channel estimation algorithm.
	
	Following the analysis in \cite{he2015convolutional}, the order of complexity for a convolutional layer and batch normalization is given by
	\begin{equation*}\label{}		
		\begin{aligned}
			\mathcal{C}_{\text{conv}} = \mathcal{O}\!\left(X Y N_I N_O S^2\right), 
		\quad 
		\mathcal{C}_{\text{BN}} = \mathcal{O}\!\left(X Y N_I\right),
		\end{aligned}	
	\end{equation*}
	\noindent where $X$ and $Y$ denote the dimensions of the output feature map, in our case, $X=N_t$ and $Y=M$ correspond to the number of BS antennas and the number of IRS elements, respectively. The term $S$ denotes the convolutional kernel size, and $N_I$ and $N_O$ represent the convolutional layer's input and output channel numbers.
	
	In addition to convolutional modules, attention mechanisms also play a role in the overall computational cost. The complexity of a single attention block can be expressed as
	\begin{equation*}\label{}		
		\begin{aligned}
			\mathcal{C}_{\text{Attention}} = \mathcal{O}\!\left(M N_t N_I\right),
		\end{aligned}	
	\end{equation*}
	\noindent where $N_I$ represents the feature dimension used for query, key, and value projections \cite{vaswani2017attention}.
	
	Combining the convolutional layers, batch normalization, and attention blocks can systematically derive the complexities of the CAN, CMN, and MB modules. Consequently, the total computational complexity of the MBA estimator across all $K$ subcarriers is given by
	\begin{equation*}\label{}		
		\begin{aligned}
			\mathcal{C}_{\text{MBA}} &= \mathcal{O}\!\Bigg(
			K\Big[
			N_t M \!\!\sum_{j \in \text{CAN}} s_j^2 n_{j-1} n_j
			+ \sum_{i=1}^{L_{\text{AB}}} M N_t N_{I,i} \\
			&+ N_t M \!\!\sum_{j \in \text{CMN}} s_j^2 n_{j-1} n_j
			+ 6 N_t M \!\!\sum_{j \in \text{MB}} s_j^2 n_{j-1} n_j
			\Big]\Bigg),
		\end{aligned}	
	\end{equation*}
	\noindent where $s_j$ denotes the convolutional kernel size of the $j^\text{th}$ layer, $n_j$ is the number of channels, $L_{\text{AB}}$ is the number of attention blocks, and $N_{I,i}$ is the input dimension of the $i^\text{th}$ attention block. Here, the notation $j \in \text{CAN}$ (or $j \in \text{CMN}$, $j \in \text{MB}$) indicates that the summation is performed over all convolutional layers belonging to the respective module.
	
	It is important to emphasize that this complexity scales linearly with $N_t$, $M$, and $K$. Unlike iterative optimization-based methods whose costs grow with the pilot length or number of iterations, the MBA method maintains a fixed per-subcarrier cost. This results in improved computational efficiency compared to state-of-the-art approaches, thereby reducing channel estimation time and making MBA especially attractive for latency-sensitive wireless communication scenarios.
	
	\section{Simulation Results}\label{sim}
	This section presents comparative simulations for MBA and state-of-the-art channel estimation methods. To have a realistic scenario, we consider 3GPP urban micro (UMi) with the CDL model.
	\subsection{Channel Model and Simulation Parameters}
	We adopt the 3GPP TR 38.901 Release 17 CDL model to simulate a mmWave channel in a UMi scenario at a carrier frequency of $f_c = 28$~GHz ~\cite{etsi_tr_138901_2022}. The BS is equipped with the ULA featuring \(N_t = 16\) transmit antennas and an IRS composed of a \(M = 12 \times 12\) UPA with passive reflecting elements. Each channel realization includes $L_{\text{IRS-BS}} = 4$ multi-path components for the IRS-BS link and $L_{\text{MU-IRS}} = 10$ for the MU-IRS link.

	The root-mean-square delay spread (DS), denoted as $\mathcal{X}_{\text{DS}}$, follows a log-normal distribution in the UMi scenario. The parameters for it are derived as follows:
	\begin{equation*}
		\begin{aligned}
			\mu_{\log_{10} \mathcal{X}_{\text{DS}}} &= -0.24 \log_{10}(1 + f_c) - 6.83, \\
			\sigma_{\log_{10} \mathcal{X}_{\text{DS}}} &= 0.16 \log_{10}(1 + f_c) + 0.28,
		\end{aligned}
	\end{equation*}
	
	\noindent where $f_c$ is expressed in GHz. Additionally, Path delays are generated as follows:
	\begin{equation*}
		\begin{aligned}
			\tau_l = -r_\tau \cdot \mathcal{X}_{\text{DS}} \cdot \log(U_l),
		\end{aligned}
	\end{equation*}
	where $r_\tau = 2.1$ and $U_l \sim \mathcal{U}(0,1)$ is a uniform random variable. Furthermore, all angular spreads depend on the frequency and have log-normal distributions, as detailed in Table 7.5-6 of the UMi scenario in~\cite{etsi_tr_138901_2022}. 

	We trained all models on the same dataset across various SNR levels. The same channel data was used to generate received pilot signals at various SNRs, which were then employed for training. The test set was generated similarly. In total, 80,000 channel samples were synthesized, with 50,000 for training, 20,000 for validation, and 10,000 for testing.
	
	We set a batch size of 64 for MBA model training and employ the Adam optimizer for both CAN and CMN, with parameters $\beta_1 = 0.9$ and $\beta_2 = 0.999$ \cite{kingma2014adam}. The CAN model adopts a learning rate scheduler that reduces the learning rate by $0.6$ every $150$ epoch, starting from $0.0002$, while the CMN model uses a fixed learning rate of $0.0001$. All deep learning-based methods, including our proposed model, were also trained using an NVIDIA Tesla T4 GPU and implemented with the PyTorch framework. 
	
	As previously mentioned, we employ the NMSE as the performance metric, defined as follows \cite{10053657}:
	\begin{equation}
		\begin{aligned}
			\text{NMSE} = \mathbb{E} \left[ \frac{\| \mathbf{H}_{\text{cs}} - \hat{\mathbf{H}}_{\text{MBA}} \|_F^2}{\| \mathbf{H}_{\text{cs}}\|_F^2} \right]
		\end{aligned}
	\end{equation}
	
	We compare the MBA against several existing estimation techniques, including LS, SMJCE \cite{10053657}, RS-OMP \cite{mao2022channel}, DD-FS \cite{9944694}, SRDnNet \cite{10025776}, DA-RLAMP \cite{10309967}, and SFCNN \cite{8752012}.
	\begin{table*}[t]
		\centering
		\caption{Comparison of NMSE performance for different activation patterns with an IRS.}
		\label{tab:pattern}
		\scalebox{0.85}{
			\begin{tabular}{c|cc|cc|cc|cc}
				\textbf{Number of Activated Elements} &  \multicolumn{2}{c|}{\textbf{Column Pattern}} &  \multicolumn{2}{c|}{\textbf{Row Pattern}}& \multicolumn{2}{c|}{\textbf{Random Pattern}} & \multicolumn{2}{c}{\textbf{Proposed Pattern}}  \\
				\cline{2-9}
				& SNR=20 dB & SNR=10 dB & SNR=20 dB & SNR=10 dB& SNR=20 dB & SNR=10 dB & SNR=20 dB & SNR=10 dB \\
				\noalign{\hrule height 1.3pt}
				$B=24$  & 0.031148 & 0.044473 & 0.029091 & 0.044093 & 0.024032 & 0.036665 & \textbf{0.011532} & \textbf{0.0166701}   \\
				\noalign{\hrule height 1pt}
				$B=48$ & 0.001802 & 0.015167 & 0.001798 & 0.015053& 0.000826 & 0.004651 & \textbf{0.000160} & \textbf{0.0014953}   \\
				\noalign{\hrule height 1.3pt}
		\end{tabular}}
	\end{table*}
		\begin{figure}[t]
		\renewcommand{\figurename}{Figure}
		\centering
		\subfloat[][]{\includegraphics[width=\linewidth]{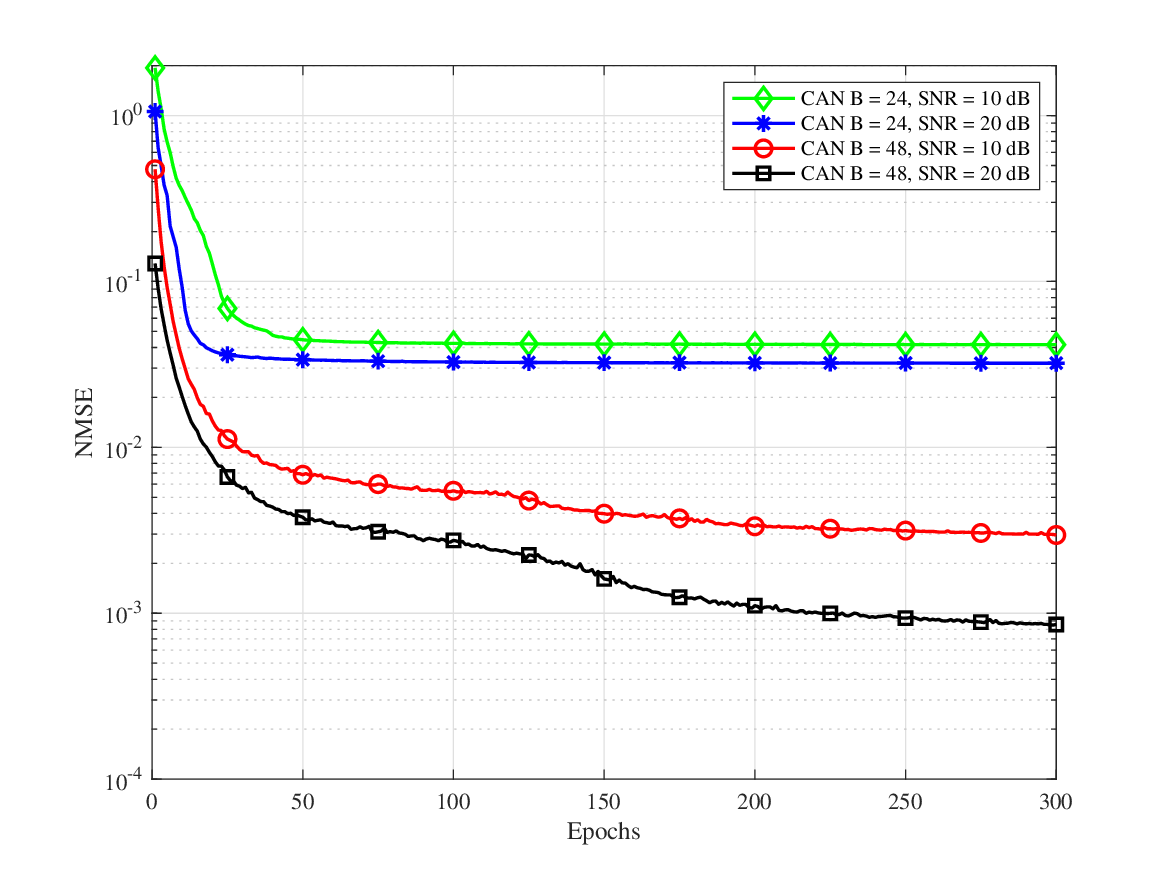}}\
		\subfloat[][]{\includegraphics[width=\linewidth]{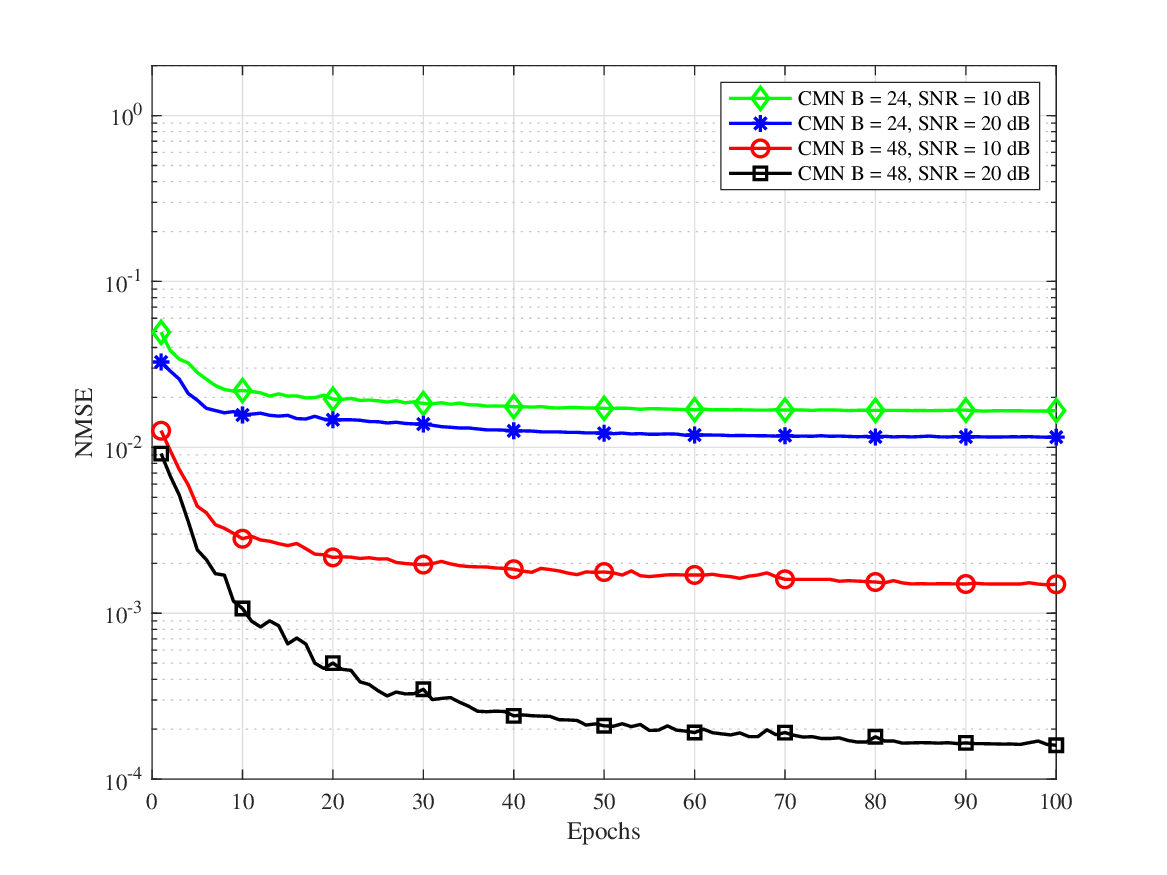}}
		\caption{ Convergence of the proposed model in terms of NMSE over epochs. (a) CAN model; (b) CMN model.}
		\label{conv}
	\end{figure}	
	\subsection{Analysis of MBA}
	We begin by analyzing the convergence behavior of the CAN and CMN models, which are the components of the MBA, separately. Figure \ref{conv} presents the NMSE convergence of the two models under SNRs of $10$ dB and $20$ dB on the test set. The CAN model converges within $100$ epochs for $B = 24$, while increasing the number of pilot signals to $B = 48$ extends the training time to approximately $300$ epochs. In contrast, the CMN model exhibits rapid and stable convergence across all cases, indicating strong robustness in noise suppression.
	
	To assess the impact of IRS activation strategies, we evaluate MBA under various IRS element deactivation patterns. As shown in Table \ref{tab:pattern}, our proposed activation scheme achieves the lowest NMSE. This is due to preserving a higher degree of spatial correlation compared to both random and structured deactivation baselines. Given these results, the proposed pattern is adopted in the rest of the simulations.
		
	As explained in the theoretical analysis in Section \ref{theory}.B, the parameters $\lambda_{\text{CAN}}$ and $\lambda_{\text{CMN}}$ characterize the feature recovery and noise suppression capabilities of CAN and CMN, respectively. Table \ref{tab:MBA_analysis} presents these values for different pilot signal counts at SNRs of $10$ dB and $20$ dB. As observed, both $\lambda_{\text{CAN}}$ and $\lambda_{\text{CMN}}$ approaches $1$ as the number of pilot signals increases, which correlates with a reduction in channel estimation error.

\begin{figure}[t]
		\renewcommand{\figurename}{Figure}
		\centering
		\includegraphics[width=\linewidth]{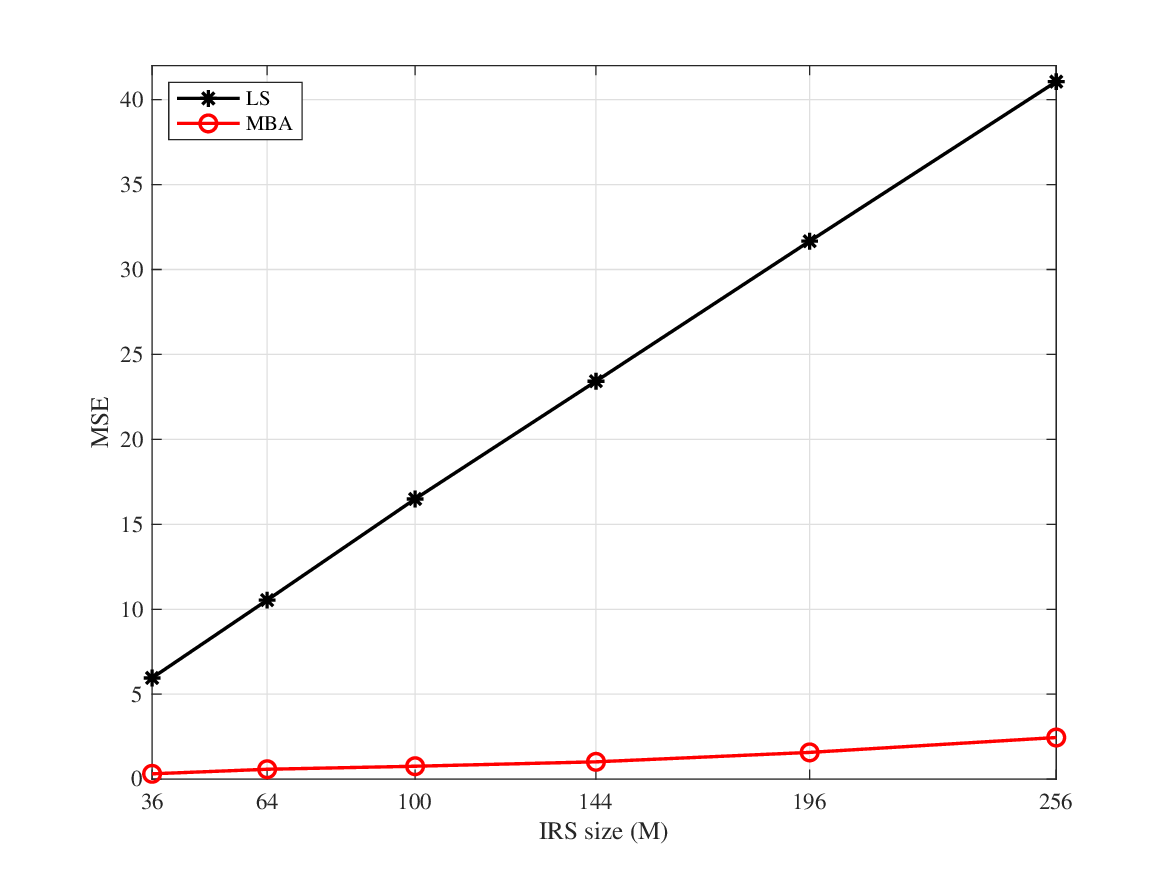}
		\caption{MSE comparison between the LS estimator and the MBA model for different IRS sizes.}
		\label{fig:j}
	\end{figure}

	Figure \ref{fig:j} illustrates the MSE performance of both the LS estimator and the proposed MBA model across different IRS sizes. For this experiment, the number of training signals for MBA is fixed at $B = 48$. The results confirm that while the MSE of the LS estimator increases linearly with IRS size, the MBA model consistently achieves lower MSE due to its superior feature recovery and denoising capabilities. This demonstrates that our model not only enhances estimation accuracy but also reduces the dependence on extensive pilot training, making it suitable for large-scale IRS-assisted systems.
	\begin{table}[t]
		\centering
		\caption{analysis of feature recovery and noise suppression gain}
		\label{tab:MBA_analysis}
		\scalebox{0.89}{
			\begin{tabular}{c|cc|cc}
				\textbf{Training Signal} &  \multicolumn{2}{c|}{$\lambda_{\text{CAN}}$} &  \multicolumn{2}{c}{$\lambda_{\text{CMN}}$}  \\
				\cline{2-5}
				& SNR=10 dB & SNR=20 dB & SNR=10 dB & SNR=20 dB \\
				\noalign{\hrule height 1.3pt}
				$B=12$  & 0.5175 & 0.5535  & 0.1980 &  0.2521    \\
				\noalign{\hrule height 1pt}
				$B=18$ & 0.8345 & 0.8541 & 0.4350 & 0.5121 \\
				\noalign{\hrule height 1.3pt}
				$B=24$  & 0.9508 & 0.9544  & 0.6028 & 0.6972    \\
				\noalign{\hrule height 1pt}
				$B=36$  & 0.9871 & 0.9905 & 0.6555 &  0.7903  \\
				\noalign{\hrule height 1pt}
				$B=48$  & 0.9950 & 0.9988  & 0.6741 & 0.8935    \\
				\noalign{\hrule height 1pt}
				$B=72$  &0.9954  & 0.9993  & 0.7072 & 0.9150    \\
				\noalign{\hrule height 1pt}
		\end{tabular}}
	\end{table}	
	\subsection{Performance Comparison of Different Estimation Schemes}
\begin{figure}[t]
		\renewcommand{\figurename}{Figure}
		\centering
		\subfloat[][]{\includegraphics[width=\linewidth]{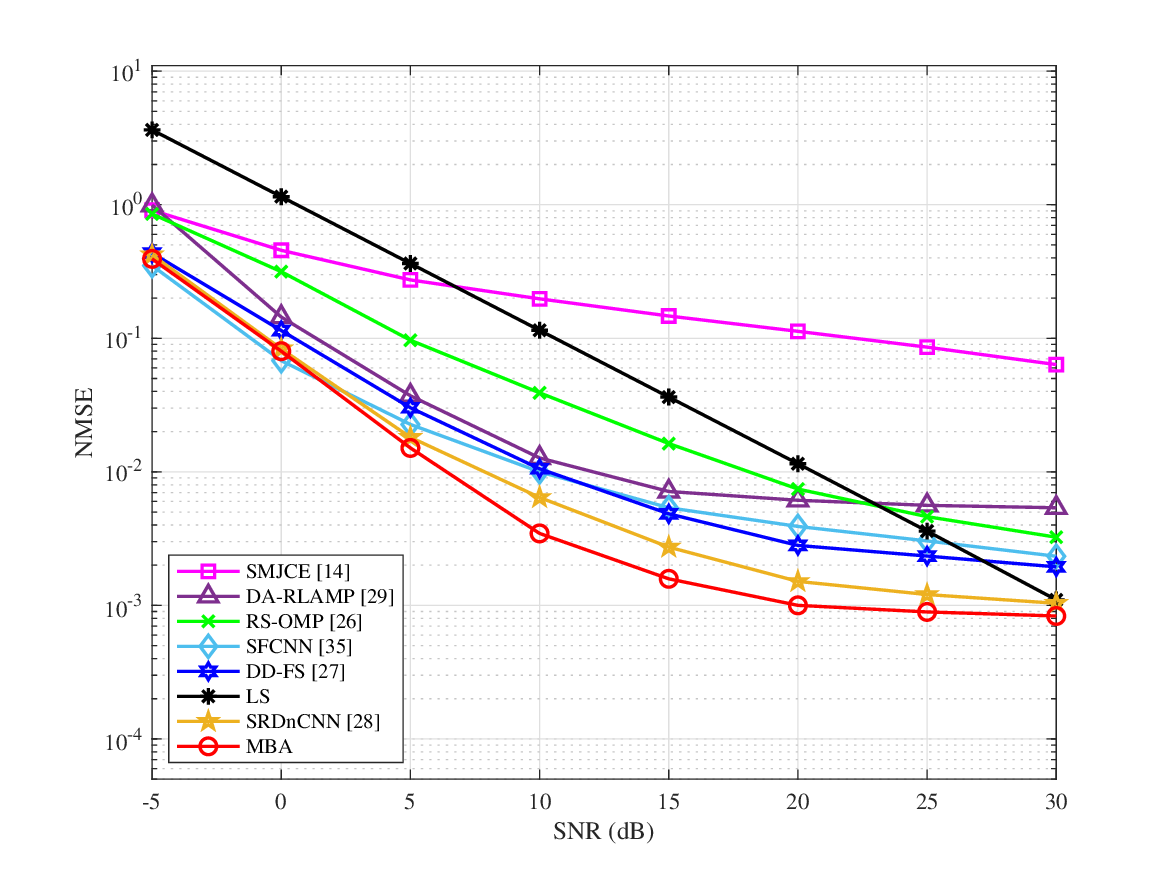}}\
		\subfloat[][]{\includegraphics[width=\linewidth]{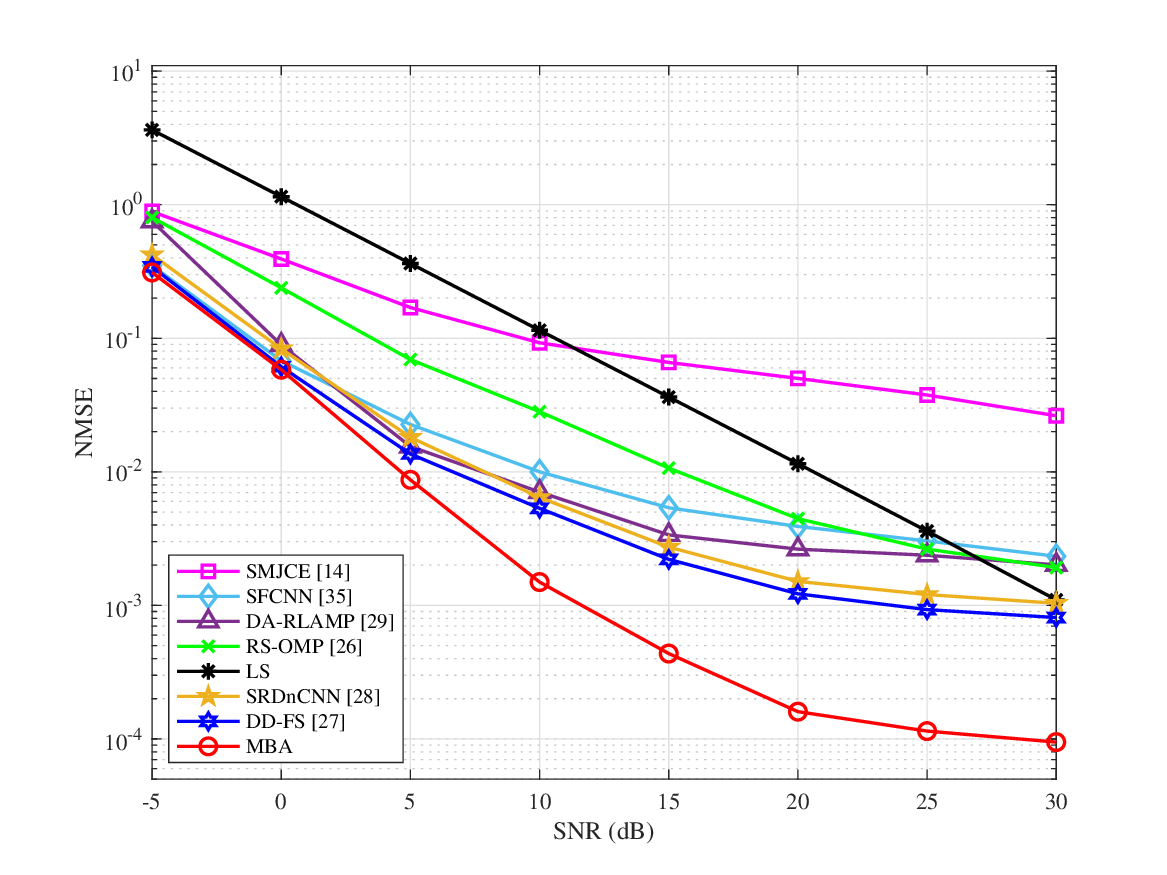}}
		\caption{NMSE comparison of different methods for $\text{N}_t = 16$ and $\text{M} = 144$: (a) $B = 36$; (b) $B = 48$.}
		\label{fig:snr}
	\end{figure}

	To evaluate the effectiveness of the proposed method, we compare the NMSE across different SNR levels with several state-of-the-art baseline methods, as shown in Fig.~\ref{fig:snr}. The evaluation considers pilot lengths of $B=36$ and $B=48$ for methods capable of operating under reduced pilot overhead.
	
	As expected, the LS estimator is highly noise-sensitive, leading to severe NMSE degradation at low SNRs. It is important to highlight that LS, SFCNN, and SRDnNet require the number of pilots $(B)$ to exceed the number of IRS elements $(M)$. Therefore, in Fig. \ref{fig:snr}, where $M = 144$, these methods do not perform properly when $B = 36$ or $48$. The corresponding performance curves for these methods are generated for $B = 144$ and are presented solely for comparative purposes. Despite the substantially greater pilot overhead associated with LS, SFCNN, and SRDnNet ($144$ as opposed to $36$ or $48$), the MBA algorithm consistently demonstrates superior performance.
	
	The RS-OMP, SRDnNet, DD-FS, and DA-RLAMP methods demonstrate stronger noise robustness than the SMJCE technique by exploiting learning-based mechanisms for effective noise suppression. In particular, the DD-FS model further enhances estimation accuracy by incorporating a dual-domain sparsity structure.
	
	As illustrated in Fig.~\ref{fig:snr}(b), increasing the number of pilot signals significantly enhances the performance of the proposed MBA model, especially at SNR levels above $10$ dB. Comparing SRDnNet and DD-FS, we observe that increasing the pilot number from $36$ to $48$ leads to a more substantial performance gain for DD-FS. This improvement is attributed to the availability of more measurements, enabling DD-FS to surpass SRDnNet under these conditions. Furthermore, the DA-RLAMP approach exhibits strong dependence on the number of BS measurements; increasing the pilot signals from $36$ to $48$ significantly improves its estimation accuracy. However, beyond approximately $15$ dB SNR, DA-RLAMP reaches a performance plateau, unlike other methods that continue to improve.
	
	\subsection{Effect of Pilot Overhead on Estimation Accuracy and Model Generalization}
\begin{figure}[t]
		\renewcommand{\figurename}{Figure}
		\centering
		\includegraphics[width=\linewidth]{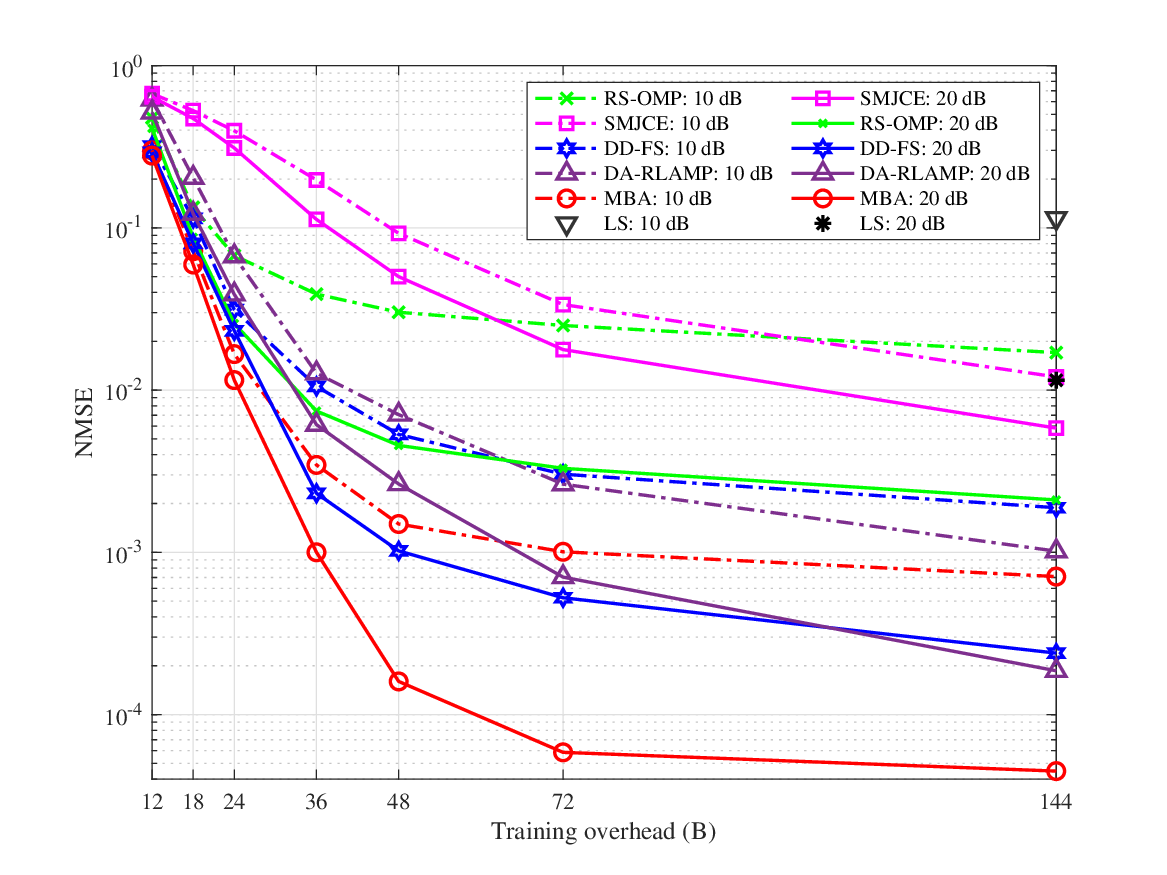}
		\caption{Impact of the number of training signals on NMSE performance with: $\text{N}_t=16$, $\text{M}=144$.}
		\label{fig:pilot}
	\end{figure}

	This subsection analyzes the impact of the number of pilot signals on channel estimation performance and evaluates the generalization capabilities of various estimation techniques, including SMJCE, RS-OMP, DD-FS, DA-RLAMP, and the proposed MBA model. Specifically, we investigate how the number of pilot signals affects the NMSE at two SNR levels: $10$ dB and $20$ dB. As shown in Fig. \ref{fig:pilot}, increasing the number of pilot signals consistently enhances estimation accuracy across all methods evaluated.
	
	The CS-based approaches, including DD-FS, RS-OMP, DA-RLAMP, and SMJCE, exhibit noticeable NMSE reductions as the number of pilots increases from $B = 72$ to $B = 144$, with SMJCE showing the most pronounced improvement due to the availability of more measurements that enhance CS recovery. In addition, the accuracy of the DA-RLAMP approach is poor with few training signals; however, the accuracy improves significantly once the number of pilots increases, with steeper NMSE reductions between $B = 48$ to $B = 72$ and $B = 72$ to $B = 144$. At an SNR of $10$ dB, the proposed MBA method achieves substantially lower NMSE, outperforming DD-FS by $51\%$, RS-OMP by $76\%$, DA-RLAMP by $62\%$, and SMJCE by $90\%$.
	
	To further assess the robustness and generalization ability of the estimation methods, we evaluate their performance across different propagation environments. Here, generalization refers to the ability of a model trained under one scenario to maintain reliable performance when applied to unseen environments. According to the 3GPP TR 38.901 specification \cite{etsi_tr_138901_2022}, the propagation characteristics, such as power delay profiles and angular spreads, vary significantly between the UMi and urban macro (UMa) scenarios. To this end, all models, including MBA, DD-FS, DA-RLAMP, and RS-OMP, are trained using channel data generated under the UMi scenario and then tested on the UMa scenario without any adaptation or prior knowledge of the UMa channel statistics.
	
	\begin{figure}[t]
		\renewcommand{\figurename}{Figure}
		\centering
		\includegraphics[width=\linewidth]{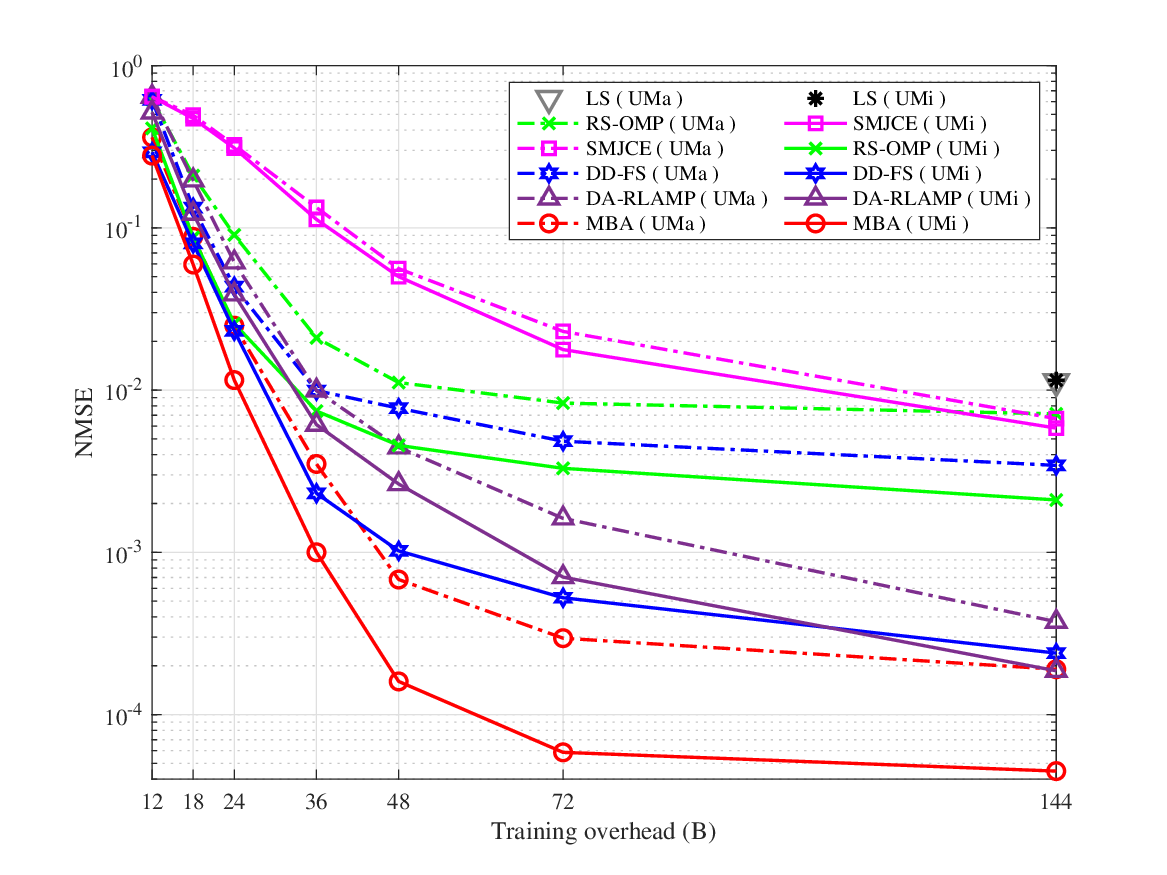}
		\caption{Generalization analysis of different models in UMa and UMi scenarios with: $\text{N}_t=16$, $\text{M}=144$.}
		\label{fig:umi_uma}
	\end{figure}	
	
	\begin{figure}[t]
		\renewcommand{\figurename}{Figure}
		\centering
		\includegraphics[width=\linewidth]{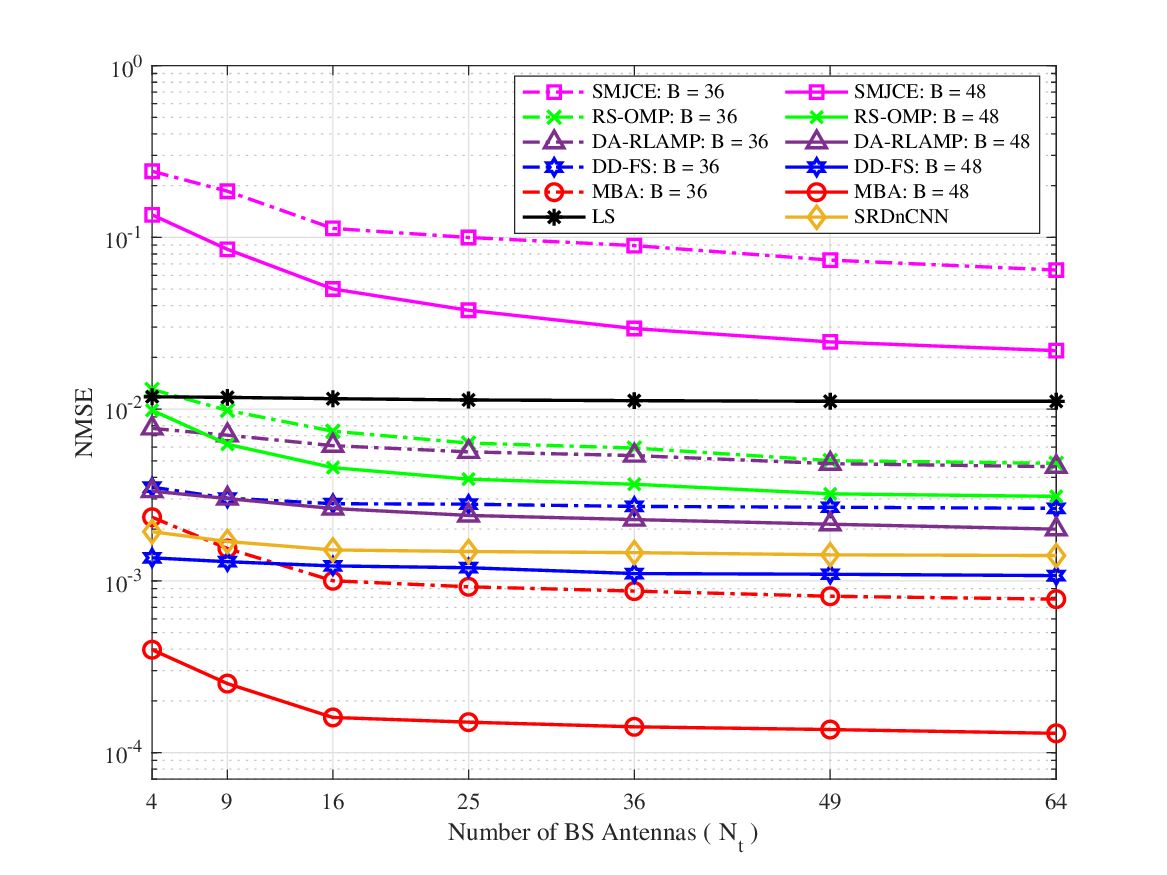}
		\caption{Effect of the number of BS antennas on NMSE performance: $\text{M} = 144$}
		\label{fig:nt}
	\end{figure}

	\begin{figure}[t]
		\renewcommand{\figurename}{Figure}
		\centering
		\includegraphics[width=\linewidth]{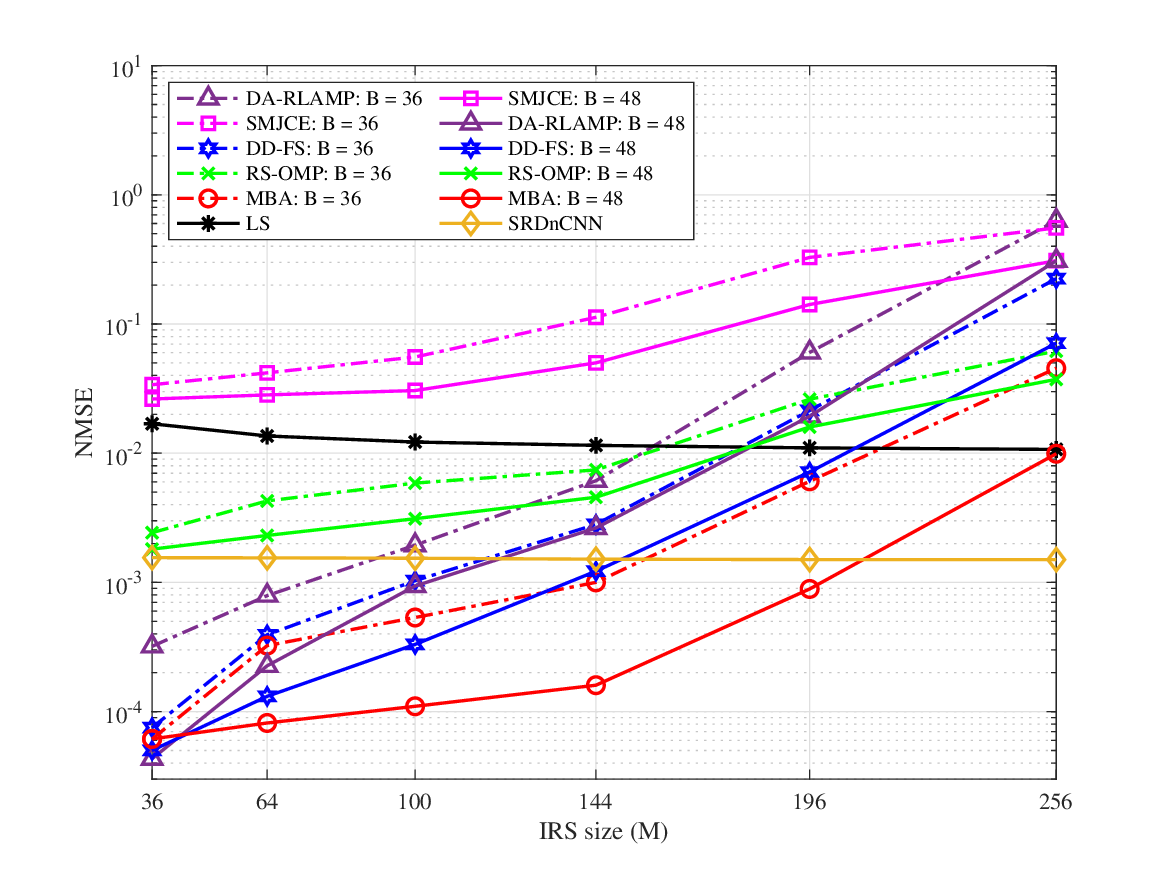}
		\caption{Impact of the number of elements in the IRS on NMSE performance: $\text{N}_t = 16$}
		\label{fig:irs}
	\end{figure}	
	
	As illustrated in Fig.~\ref{fig:umi_uma}, the proposed MBA model and DA-RLAMP demonstrate superior generalization performance. Since RS-OMP employs an initial channel estimation stage using CS-based algorithms to generate inputs for its DNN, it also achieves a reasonable level of generalization. In contrast, the DD-FS method suffers from more severe performance degradation than RS-OMP. Similar to RS-OMP, DA-RLAMP leverages a pre-estimation stage, enabling its DNN model to generalize well across different scenarios. Nevertheless, the estimation accuracy of DA-RLAMP remains considerably lower than that of the proposed MBA, with reductions of approximately $73\%$ in the UMi scenario and about $62\%$ in the UMa scenario.
	
	\subsection{Effect of the Number of BS Antennas and IRS Size on Channel Estimation Performance}
	Figure \ref{fig:nt} illustrates the impact of the number of BS antennas on NMSE under two pilot overhead settings: $B = 36$ and $B = 48$, with the SNR fixed at $20$ dB. The results show that increasing the number of BS antennas has a limited effect on estimation accuracy. A modest improvement is observed when the number of antennas increases from $4$ to $16$, particularly for DL-based methods. This is likely because smaller input dimensions limit the effectiveness of feature extraction. Overall, these results suggest that the IRS size has a more pronounced influence on cascaded channel estimation performance than the number of BS antennas.
	
	Figure \ref{fig:irs} presents the effect of IRS size on NMSE performance under the same pilot overhead settings and SNR. As the number of IRS elements increases, more pilot signals are required to preserve estimation accuracy. This trend reveals a key limitation of CS-based methods, where larger IRS sizes necessitate proportional dictionary expansion, significantly increasing computational complexity.

	A comparison between DD-FS and the proposed MBA model shows that DD-FS and DA-RLAMP slightly outperform MBA when the IRS size is $B = 48$. This observation is consistent with the fact that DD-FS performs better when the number of pilot signals exceeds the number of IRS elements. However, as the IRS size increases or the number of pilots decreases, the performance of DD-FS and DA-RLAMP degrades significantly, revealing its limited scalability and adaptability. This degradation is even more pronounced in DA-RLAMP, particularly for larger IRS sizes.

	In contrast, the NMSE performance of LS and SRDnNet remains relatively stable as the IRS size increases. This indicates that these methods implicitly distribute pilot resources proportionately to the IRS size, maintaining consistent performance across different configurations.
	
	\subsection{Computational Complexity and Processing Time Analysis}
	Computational complexity plays a critical role in channel estimation, directly impacting estimation latency and the feasibility of real-time deployment. Table~\ref{tab:clOrder} presents a comparative analysis of the computational complexity for the proposed MBA model and several benchmark methods. Notably, the complexity of MBA, as well as that of SFCNN, DA-RLAMP, and SRDnCNN, scales linearly with the number of BS antennas ($N_t$), the number of IRS elements ($M$), and the number of subcarriers ($K$), making them well-suited for large-scale deployments. In contrast, methods such as RS-OMP, DD-FS, and SMJCE exhibit computational burdens that additionally depend on the number of pilot symbols ($B$), resulting in significantly higher complexity under more intensive training conditions.
	\begin{table}[t]
		\begin{center}
			\caption{Computational complexity order of different approaches.}
			\label{tab:clOrder}
			\scalebox{0.815}{
				\begin{tabular}{ c | c | c}
					\textbf{Method} & \textbf{Complexity Order} & \textbf{Estimation Time}\\
					\noalign{\hrule height 1.5pt}
					MBA &  $\mathcal{C}_{\text{MBA}} \approx \mathcal{O}(2.5\times 10^5  N_tMK)$ &  $41.77$ ms \\
					\hline		
					RS-OMP & $\mathcal{C}_{\text{RS-OMP}} \approx \mathcal{O}((4\times10^5  N_tM + N_tMB)K)$ & $281.12$ ms \\
					\hline 			
					SMJCE & $\mathcal{C}_{\text{SMJCE}} \approx \mathcal{O}(((M^2(L_{\text{BS-IRS}}+2B))K)$& $2512.32$ ms \\
					\hline 
					SRDnNet & $\mathcal{C}_{\text{SRDnNet}} \approx \mathcal{O}(13.7\times10^5N_tMK)$ &  $55.43$ ms \\
					\hline
					DA-RLAMP & $\mathcal{C}_{\text{DA-RLAMP}} \approx \mathcal{O}(N_tM\times(48BK + 9.5\times10^6))$ &  $63.87$ ms \\
					\hline
					SFCNN & $\mathcal{C}_{\text{SFCNN}} \approx \mathcal{O}(3\times10^5 N_tMK)$ &  $10.54$ ms \\
					\hline
					DD-FS & $\mathcal{C}_{\text{DD-FS}} = \mathcal{O}(B(N_t+M)+2.6\times 10^5 N_tMK)$  & $66.43$ ms\\
					\hline         
					LS & $\mathcal{C}_{\text{LS}} = \mathcal{O}(  N_tM^2K)$ &  $16.59$ ms  \\
					\noalign{\hrule height 1.5pt}				
			\end{tabular}}
		\end{center}
	\end{table}	
	
	To assess the practical impact of computational complexity, we fixed the pilot overhead at $B = 48$ and measured the average estimation time over 10,000 test samples across all subcarriers. The proposed MBA framework achieves significant computational savings, reducing runtime by about $85\%$ compared to RS-OMP, $37\%$ compared to DD-FS, $34\%$ compared to DA-RLAMP, and $24\%$ compared to SRDnCNN, while still delivering higher estimation accuracy than all fast baseline methods. Although SFCNN and LS achieve faster processing, their practicality is limited since they require pilot overhead equal to or larger than the number of IRS elements.
	
	\section{Conclusion}\label{conc}
	This work introduces a deep learning-based framework for efficient cascaded channel estimation in IRS-assisted mmWave MIMO systems. We propose an IRS element deactivation strategy that selectively reduces training dimensionality while preserving spatial coherence to address the significant pilot overhead associated with large IRS deployments. We analytically demonstrate that DFT and Hadamard matrices provide optimal phase configurations for minimizing estimation error in LS estimation. To mitigate the challenges of spatial decorrelation and noise amplification arising from deactivation, we develop the MBA architecture comprising two modules: the CAN for feature restoration and the CMN for denoising. The MBA framework is analytically shown to suppress error propagation and exhibits linear complexity with respect to IRS and BS dimensions. Extensive simulations conducted under standardized mmWave channel models demonstrate that the proposed method can estimate the cascade channel approximately $24\%$ faster than state-of-the-art methods due to its lower complexity. Additionally, it achieves up to $87\%$ less pilot overhead compared to LS estimation. Furthermore, the MBA model effectively generalizes to different deployment scenarios, confirming its scalability and practical viability. Future research will focus on extending this work to multi-IRS and distributed IRS architectures, aiming for more dynamic and flexible configurations.

	\bibliographystyle{IEEEtran}
	\bibliography{reference}
	
\end{document}